\documentclass{jfm}

\begin{document}

\newtheorem{lemma}{Lemma}
\newtheorem{corollary}{Corollary}

\shorttitle{Transitions near the onset of low Prandtl-number rotating magnetoconvection} %for header on odd pages
\shortauthor{Ghosh et al.} %for header on even pages

\title{Transitions near the onset of low Prandtl-number rotating magnetoconvection}

\author
 {
 Manojit Ghosh\aff{1},
%  \corresp{\email{jfm@damtp.cam.ac.uk}},
  Paromita Ghosh\aff{1},
  Yada Nandukumar\aff{2}
  \and
  Pinaki Pal\aff{1}
  \corresp{\email{pinaki.pal@maths.nitdgp.ac.in}}
  }

\affiliation
{
\aff{1}
Department of Mathematics, National Institute of Technology, Durgapur~713209, India
\aff{2}
Centre for Theoretical Studies, Indian Institute of Technology, Kharagpur~721302, India
}

\maketitle

\begin{abstract}
We investigate the onset of convective instability and subsequent transitions near it in Rayleigh-B\'{e}nard convection (RBC) of electrically conducting low Prandtl-number ($\mathrm{Pr}$) fluids in simultaneous presence of rotation about vertical axis and external uniform horizontal magnetic field with free-slip boundary conditions by performing three dimensional direct numerical simulations (DNS). DNS shows various stationary as well as time dependent patterns including 2D rolls, cross rolls, oscillatory cross rolls at the onset of convection in presence of weak rotation and magnetic field which is intriguing. Most interestingly, the Nusselt number ($Nu$) computed from the DNS data of $2D$ rolls solutions shows a sharp jump at the onset, indicating substantial enhancement of the convective heat transfer. To understand the origin of these flow patterns as well as the reason for notable increase of heat transfer near the onset of convection, we construct low dimensional models from DNS data. Bifurcation analysis of these models help to understand the bifurcation structure and the origin of different flow patterns near the onset of convection quite convincingly in a range of Chandrasekhar number ($0<\mathrm{Q}\leq 50$) and Taylor number ($0<\mathrm{Ta}\leq 100$). Simultaneous performance of DNS and low dimensional modeling establishes a first order transition to convection at the onset in a wide range of the parameter space which is responsible for the enhancement of the heat transfer at the onset. The first order transition is found to be associated with a subcritical pitchfork bifurcation of the conduction branch at the onset of convection. Moreover, a nonstandard subcritical pitchfork bifurcation of the conduction branch occurs at a supercritical Rayleigh number ($\mathrm{Ra}$) which adds richness to the dynamics near the onset of convection. Further, in a range of $\mathrm{Ra}$, we observe imperfect gluing bifurcation in this system and explore it in detail.
\end{abstract}

\section{Introduction}
The convection of a layer of electrically conducting fluid confined between two horizontal conducting plates and heated from below (classically known as Rayleigh-B\'{e}nard convection (RBC)) in simultaneous presence of external uniform magnetic field 
and rotation provides a simplified model of convection for geophysical and astrophysical situations~\citep[]{chandra:book, Eltayeb:1972, Soward:1980, Olson:2001, aujogue:2015}. Mathematical description of RBC in presence of external magnetic field and rotation (rotating magnetoconvection (RMC)) requires five dimensionless parameters namely the Rayleigh-number ($\mathrm{Ra}$, measures the vigor of buoyancy force), the Chandrasekhar number ($\mathrm{Q}$, measures the strength of magnetic field), the Taylor number ($\mathrm{Ta}$, measures the rotation rate), the Prandtl-number ($\mathrm{Pr}$, the ratio of kinematic viscosity and thermal diffusivity of the fluid) and the magnetic Prandtl-number ($\mathrm{Pm}$, the ratio of kinematic viscosity and magnetic diffusivity of the fluid).  

RBC in absence of rotation and magnetic field has been studied for more than a century to investigate basic features of convection including instabilities, bifurcations, heat transfer, transition to turbulence etc. and still is an active area of research~\citep[]{busse:1978, peltierw:book_1989, busse:1989, bodenschatz:ARFM_2000, marshall:1999, hartman:2001, spiegel:JGR_1962,  swinney_gollub:book_1985,ahlers:RMP_2009,lohse_xia:ARFM2010}. Numerous theoretical as well as experimental studies on RBC not only have enhanced the understanding of the basic physics of convection of different fluids but also contributed substantially to the development of the subjects like hydrodynamics instabilities~\citep[]{chandra:book, drazin:book}, nonlinear dynamics~\citep[]{mannevile:book_1990, getling:book, busse:1978}, and pattern formation~\citep[]{croquette1:Contemp.Phys_1989, croquette2:Contemp.Phys_1989, cross:RMP_1993}. It has been found that the onset of stationary convective instability is independent of $\mathrm{Pr}$ but subsequent transitions as the value of $\mathrm{Ra}$ is increased, depend on it. Interesting to note here that for low Prandtl number fluids, due to the dominance of the velocity nonlinearity in the momentum equation over the nonlinearity in the energy balance equation, vertical vorticity is generated very close to the onset of convection~\citep[]{busse:JFM_1972} and as a result, rich bifurcation structures are observed there~\citep[]{schluter:1965, willis_deardorff:JFM_1970, mauer_libchaber:JPL_1980, coste:PLA84_1981, lyubimov:physicaD_1983, jenkins_proctor:JFM_1984, meneguzzi:1987, chiffaudel_et.al:EPL_1987, fauve_et.al:PhysicaD_1987, clever:POF_1990, busse:PhysicaD_1992, hof_et.al:JFM_2004, mishra:EPL_2010, pal:PRE_2013, dan:2014, dan:2015, nandu:2016}. Also RBC of electrically conducting low Prandtl number fluids has been investigated experimentally as well as theoretically in presence of horizontal~\citep[]{libchaber_et.al:JPL_1982, busse:1983, fauve:1984, fauve_prl:1984, muller:JFM_2002, yanagisawa_2010, pal:2012, nandu:2015} and vertical~\citep[]{nakagawa:1955, busse:1982, busse:1989a, cattaneo:astrophys_J_2003, arnab:2014, arnab:2015, arnab:2016} external magnetic fields. It has been revealed that horizontal magnetic field does not influence the onset of instability but vertical magnetic field inhibits the onset of convection~\citep[]{nakagawa:1955}. Both theoretical and experimental works on RBC in presence of horizontal as well as vertical magnetic fields reveal the stabilizing effects of magnetic field on convective instabilities~\citep[]{busse:1982, busse:1983, fauve:1984}. Stronger external magnetic (horizontal or vertical) fields induce two dimensionality in the convective flow~\citep[]{busse:1982, busse:1983}. Researchers further investigated RBC in presence of rotation on convective flow in detail~\citep[]{nakagawa_a:1955, veronis:1959, veronis:1966, rossby:JFM_1969, riahi:1992, cardin:1994, hu:1997, bajaj:1998, knobloch:1998, clever:2000, dawes:2000, cox:2000, prosperetti:2012, maity:EPL_2013, hirdesh:2013, knobloch:2013, kaplan:2017}. It has been established that rotation about a vertical axis inhibits the onset of convection and the RBC system shows nontrivial effects of rotation on convective instabilities~\citep[]{nakagawa_a:1955} including subcriticality~\citep[]{veronis:1966, knobloch:2013, kaplan:2017}.  

On the other hand, literature on RMC with Rayleigh-B\'{e}nard geometry is limited. The problem has been investigated by Chandrasekhar~\citep[]{chandra:book} for the first time by performing linear stability analysis of the conduction state in presence of vertical magnetic field and rotation about vertical axis with free-slip, conducting and insulating boundary conditions for velocity, temperature and magnetic fields respectively. From linear theory, Chandrasekhar determined critical Rayleigh number ($\mathrm{Ra_c}$) and wave number ($\mathrm{k_c}$) at the onset of convection for wide ranges of $\mathrm{Ta}$ and  $\mathrm{Q}$. Theoretical findings of Chandrasekhar have been experimentally verified in~\citep[]{nakagawa:1957, nakagawa:1959}. Subsequently, in a detailed theoretical analysis (based on linear theory) of the RBC system in presence of external magnetic field of various orientations, rotation about different axis and different types of boundary conditions, Eltayeb determined a list of scaling laws for the onset of convection in the infinite Taylor number and Chandrasekhar number limits~\citep[]{Eltayeb:1972}.
Similar RMC systems consisting of electrically conducting inviscid fluid in presence of horizontal magnetic field and rotation about vertical axis have been theoretically investigated for free-slip and no-slip boundary conditions in~\citep[]{Roberts:1975} and~\citep[]{Soward:1980} respectively. Both the works reported the existence oblique rolls and theoretically investigated the stability of these rolls. For large Prandtl-number electrically conducting fluids, the RMC system in presence of horizontal magnetic field and rotation about vertical axis have been theoretically investigated in~\citep[]{Roberts:2000, Jones:2000} to determine the preferred mode of convection at the onset. Aurnou and Olson~\citep[]{Olson:2001} performed experiment with low Prandtl-number electrically conducting fluid liquid gallium in presence of vertical magnetic filed and rotation about vertical axis and found that heat transfer is inhibited for a high Taylor number which is corroborated in a theoretical investigation by Zhang et al.~\citep[]{zhang:2004}. A subsequent numerical study~\citep[]{Varshney:2008, Baig:2008} investigated RMC both in presence of horizontal and vertical magnetic fields. They studied the heat transfer properties of the fluid. Later~\citep[]{Podvigina:2010, Eltayeb:2013} theoretically studied RMC for various boundary conditions in presence of rotation and magnetic field determined the preferred mode of convection at the onset of instability. In a recent study~\citep{ghosh:2017}, RMC in presence of rotation about the vertical axis and horizontal magnetic field has been investigated numerically in the asymptotic zero Prandtl-number limit. They discovered an interesting bifurcation structure including subcritical bifurcation near the onset of convection. However, the transition to convection and subsequent bifurcations along with heat transfer properties near the onset of stationary convection have not been investigated in RMC of low Prandtl-number electrically conducting fluids.  

In this paper, we perform direct numerical simulations (DNS) of the RMC system near the onset of stationary convection of low Prandtl-number fluids in presence of rotation about vertical axis and horizontal uniform magnetic field using random initial conditions.  DNS of the system shows a variety of stationary as well as time-dependent convective patterns just at the onset of convection including finite amplitude 2D rolls, cross rolls, oscillatory cross rolls for different values of the governing parameters. Moreover, the Nusselt number (measure of convective heat transfer) computed from the DNS data at the onset shows a significant jump. Inspired by these intriguing DNS results, the present study is centered around two fundamental questions: (1) What are the origin of different flow patterns at the onset? and (2) Which causes the enhancement of heat transfer at the onset? Bifurcation analysis of the system using a couple of low dimensional models derived from DNS data is found to provide a satisfactory answer to these questions. In the subsequent sections,  we discuss the details of the analysis along with a description of the rotating hydromagnetic system. 

\section{Rotating Hydromagnetic System}
\begin{figure}
\begin{center}
\includegraphics[height=2.5cm,width=0.8\textwidth]{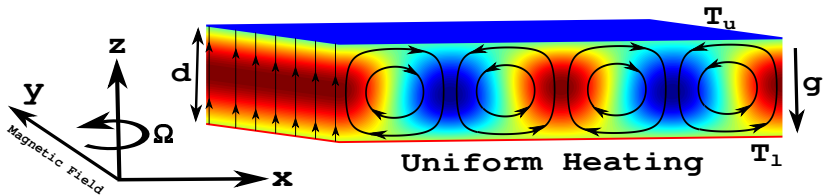}
\end{center}
\caption{Schematic diagram of rotating magnetoconvection. }\label{fig:rbc}
\end{figure}

The hydromagnetic system under consideration consists of a thin horizontal layer of electrically conducting Boussinesq fluid of thickness $d$, thermal diffusivity $\kappa$, magnetic diffusivity $\lambda$, coefficient of volume expansion $\alpha$ and kinematic viscosity $\nu$ kept between two horizontal thermally and electrically conducting plates in presence of an external uniform horizontal magnetic field ${\bf B}_0 \equiv (0, B_0, 0)$. The system rotates uniformly with angular velocity $\Omega$ about the vertical axis (along opposite direction of the gravity) and heated uniformly from below to maintain an adverse temperature gradient $\beta = \Delta \mathrm{T}/d = (\mathrm{T}_l - \mathrm{T}_u)/d$ across the fluid layer, where $\mathrm{T}_l$ and $\mathrm{T}_u$ are the temperatures of the lower and upper plates respectively and $\mathrm{T}_l > \mathrm{T}_u$. A schematic diagram of the described system has been shown in figure~\ref{fig:rbc}. The system under Boussinesq approximation is governed by the set of dimensionless equations
\begin{eqnarray}
\frac{\partial \bf{v}}{\partial t} + (\bf{v}.\nabla)\bf{v} &=& -\nabla{\pi} + \nabla^2{\bf{v}} + \mathrm{Ra} \theta {\bf{\hat{e_3}}}+ \sqrt{\mathrm{Ta}}(\bf{v}\times {\bf{\hat{e_3}}})\nonumber \\ && +\mathrm{Q}\left[\frac{\partial{\bf b}}{\partial y} + \mathrm{Pm}({\bf b}{\cdot}\nabla){\bf b}\right] , \label{eq:momentum} \\   
\mathrm{Pm}[\frac{\partial{\bf b}}{\partial t} + ({\bf v}{\cdot}\nabla){\bf b} &-& ({\bf b}{\cdot}\nabla){\bf v}] = {\nabla}^2 {\bf b} + \frac{\partial{\bf v}}{\partial y},\label{eq:magnetic}\\      
%\nabla^2\bf{b} &=& -\frac{\partial \bf{v}}{\partial y}, \label{magnetic} \\
\mathrm{Pr}\left[\frac{\partial \theta}{\partial t}+(\bf{v}.\nabla)\theta\right] &=& v_3+\nabla^2\theta \label{eq:heat},\\
\nabla . \bf{v}=0, && \nabla . \bf{b}=0, \label{eq:div_free}
\end{eqnarray}
where $\mbox{\bf {v}} (x,y,z,t) \equiv (\mathrm{v_1}, \mathrm{v_2}, \mathrm{v_3})$ is the velocity field, $\pi$ is the modified pressure field, $\mbox{\bf{b}} (x,y,z,t) \equiv (\mathrm{b_1},\mathrm{b_2}, \mathrm{b_3})$ is the induced magnetic field due to convection, $\theta (x, y, z, t)$ is the deviation in temperature field from the steady conduction profile, $g$ is the acceleration due to gravity and $\mbox{\boldmath $\hat{e}$}_3$ is  the unit vector directed in the positive direction of $z$ axis. We measure length, time, velocity, temperature field and induced magnetic field in the units of fluid thickness $d$, the viscous diffusion time $\frac{d^2}{\nu}$, $\frac{\nu}{d}$, $\frac{\beta d \nu}{\kappa}$ and $\frac{B_0\nu}{\lambda}$, respectively. Therefore the dynamics of the system is controlled by five dimensionless parameters namely the Rayleigh number $\mathrm{Ra}=\frac{\alpha \beta g d^4}{\nu \kappa}$, the Taylor number $\mathrm{Ta} = \frac{4\Omega^2d^4}{\nu^2}$, the Chandrasekhar number $\mathrm{Q}=\frac{{B_0}^2 d^2}{\nu \lambda \rho_0}$, the magnetic Prandtl number $\mathrm{Pm}=\frac{\nu}{\lambda}$ and the Prandtl number $\mathrm{Pr}=\frac{\nu}{\kappa}$. In this paper, we are interested to investigate the instabilities and bifurcations occurring near the onset of convection of electrically conducting low Prandtl number fluids ($\mathrm{Pr}\approx 10^{-2}$) for which the magnetic Prandtl number is very low ($\mathrm{Pm}\approx 10^{-6}$) and for simplicity we consider $\mathrm{Pm}\rightarrow 0$. In this limit, the equations~(\ref{eq:momentum}) and~(\ref{eq:magnetic}) reduce to
\begin{eqnarray} 
\frac{\partial \bf{v}}{\partial t} + (\bf{v}.\nabla)\bf{v} &=& -\nabla{\pi} + \nabla^2{\bf{v}} + \mathrm{Ra} \theta {\bf{\hat{e_3}}}\nonumber \\ && + \sqrt{\mathrm{Ta}}(\bf{v}\times {\bf{\hat{e_3}}})+\mathrm{Q}\frac{\partial{\bf b}}{\partial y}, \label{eq:momentum1} \\   {\mathrm{and}~~~}
\nabla^2\bf{b} &=& -\frac{\partial \bf{v}}{\partial y}. \label{eq:magnetic1} 
%\mathrm{Pr}\left[\frac{\partial \theta}{\partial t}+(\bf{v}.\nabla)\theta\right] &=& v_3+\nabla^2\theta \label{eq:heat1},\\ 
%\nabla . \bf{v}=0 &&\mathrm{and} ~\nabla . \bf{b}=0. \label{eq:div_free1}
\end{eqnarray}
%Therefore, the equations~(\ref{eq:momentum1}),~(\ref{eq:magnetic1}),~(\ref{eq:heat}) and~(\ref{eq:div_free}) constitute the set of governing equations for the present study.
The top and bottom boundaries are assumed to be perfectly thermally conducting and stress-free, which imply
%This kind of boundary conditions can be approximated by the interface between two immiscible fluids with large viscosity difference and more reliable in astrophysical problems than the laboratory experiments. Though Goldstein and Graham were successful to achieve almost perfect stress-free boundaries in their experimental work in the laboratory. 
%Therefore we have
\begin{equation}
v_3 = \frac{\partial v_1}{\partial z} = \frac{\partial v_2}{\partial z} = \theta = 0~\mathrm{at}~ z = 0, 1.\label{bc1}
\end{equation}
Periodic boundary conditions are assumed along horizontal directions. We also consider horizontal boundaries to be perfectly electrically conducting, this leads to the conditions
\begin{equation}
b_3 = \frac{\partial b_1}{\partial z} = \frac{\partial b_2}{\partial z} = 0~\mathrm{at}~ z = 0, 1.\label{bc2}
\end{equation}
Therefore, the mathematical description of the problem consists of the equations ~(\ref{eq:heat})-~(\ref{eq:magnetic1}) along with the boundary conditions (\ref{bc1}) and (\ref{bc2}). We now perform linear stability analysis of the conduction state of the system following Chandrasekhar\citep{chandra:book} and find that the onset of stationary convection is independent of $\mathrm{Pr}$ and $\mathrm{Q}$ but depends only on $\mathrm{Ta}$ in the ranges of $\mathrm{Ta}$ and $\mathrm{Q}$ we have considered in this paper. The details of the linear analysis for the present problem is same as the one presented in~\cite{ghosh:2017} and we determine critical Rayleigh number $\mathrm{Ra_c}$ as well as wave number $\mathrm{k_c}$ at the onset of stationary convection as a function of $\mathrm{Ta}$ from the analysis. 

\section{Direct numerical simulations (DNS)}
We simulate the system (\ref{eq:heat}) - (\ref{eq:magnetic1}) together with the boundary conditions (\ref{bc1}) and (\ref{bc2}) using an object oriented pseudo-spectral code Tarang \citep[]{mkv:code}. In the simulation code, independent fields are expanded either by using the set of orthogonal basis functions \{$e^{i(lk_xx + mk_yy)}\sin(n\pi z): l, m, n = 0, 1, 2,...$\} or by \{$e^{i(lk_xx + mk_yy)}\cos(n\pi z): l, m, n = 0, 1, 2,...$\}, whichever is compatible with the boundary conditions. The wavenumbers along $x$-axis and $y$-axis are denoted by $k_x$ and $k_y$ respectively. Therefore, the independent fields, vertical velocity, vertical vorticity and temperature are expanded as 
\begin{eqnarray}
v_3 (x,y,z,t)&=& \sum_{l,m,n} W_{lmn}(t)e^{i(lk_xx+mk_yy)}\sin{(n\pi z)},\nonumber\\
\omega_3 (x,y,z,t) &=& \sum_{l,m,n} Z_{lmn}(t)e^{i(lk_xx+mk_yy)}\cos{(n\pi z)},\nonumber\\
\theta (x,y,z,t) &=& \sum_{l,m,n} T_{lmn}(t)e^{i(lk_xx+mk_yy)}\sin{(n\pi z)},\nonumber\\
\end{eqnarray}
where $W_{lmn}$, $Z_{lmn}$ and $T_{lmn}$ are Fourier amplitudes or modes. Components of horizontal velocity and vorticity are derived using the continuity equation. Induced magnetic field is determined from equation (\ref{eq:magnetic1}). For the present simulation we take $k_x = k_y = k_c$, the critical wave number determined from the linear theory~\citep[]{ghosh:2017}. Simulations are performed using random initial conditions in a square simulation box of size $(2\pi/k_c) \times (2\pi/k_c) \times 1$. We have taken $32^3$ spatial grid resolution for the simulations. Fourth order Runge-Kutta (RK4) integration scheme with Courant-Friedrichs-Lewy (CFL) condition is used for time advancement with time step $\delta t = 0.001$. In the subsequent discussion we have used a parameter $r$, called the reduced Rayleigh number defined by $r = \mathrm{Ra}/\mathrm{Ra_c}$.

%Therefore, we started simulation for a given set of values of $\mathrm{Ta}$, $\mathrm{Q}$ and $\mathrm{r}$ and increased the value of $r$ in small steps. We also performed simulations for higher values of $r$ and decreased its value by small steps by using the final values of all the fields of the last simulation as the current initial conditions. 
%\\
\begin{figure}
\begin{center}
\includegraphics[height=!, width=\textwidth]{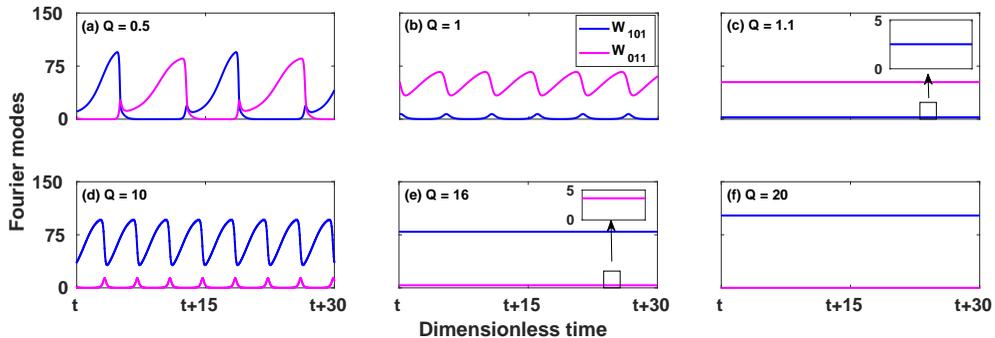}
\end{center}
\caption{Temporal variation of two largest Fourier modes $W_{101}$ (blue curves) and $W_{011}$ (pink curves) near the onset of convection ($r = 1.001$) for $\mathrm{Ta} = 20$ and different values of $\mathrm{Q}$ (a)-(f) showing different flow regimes at the onset.}
\label{fig:tau_20_diff_Q_onset}
\end{figure}
\begin{figure}
\begin{center}
\includegraphics[height=!, width=\textwidth]{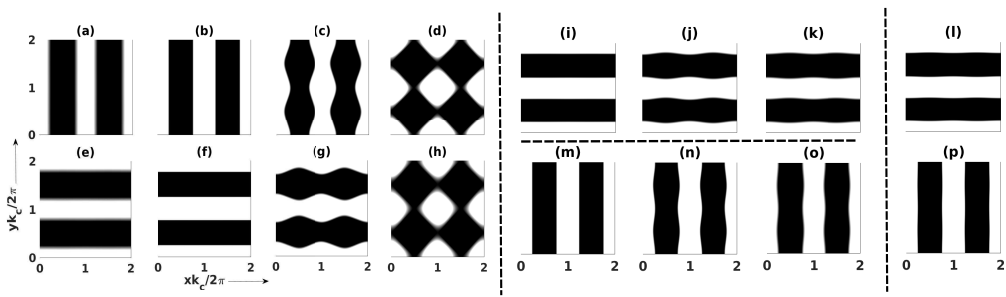}
\end{center}
\caption{Isotherms computed from DNS at $z = 0.5$ near the onset of convection ($r = 1.001$) showing pattern dynamics of different solutions. Fig (a)-(h) display pattern dynamics of STO. Pattern dynamics corresponding to oscillatory cross rolls is shown (i)-(k) (OCR-II$'$) and (m)-(o) (OCR-II). Whereas fig (l) and (p) display the pattern dynamics of CR$'$ and CR solutions respectively.}
\label{fig:dns_pattern}
\end{figure}
\section{Results and Discussions}
\subsection{DNS Results}
We now perform DNS of the system in presence of rotation and magnetic field for $\mathrm{Pr} = 0.025$ near the onset of convection.  First, we set $\mathrm{Ta} = 20$ and slowly increase the value of $\mathrm{Q}$ starting from a small value. Figure~\ref{fig:tau_20_diff_Q_onset} shows the temporal evolution of the Fourier modes $W_{101}$ and $W_{011}$ as obtained from DNS for different values of $\mathrm{Q}$. From figure~\ref{fig:tau_20_diff_Q_onset}(a) we observe that for very small value of $\mathrm{Q}$, the Fourier mode $W_{101}$ grows exponentially and then rapidly decays to zero in a very short period. The mode $W_{011}$ is excited from zero shortly before the mode $W_{101}$ reaches its maximum. However, once either of the modes reaches zero, it spends significant time there before bursting to a large amplitude. This type of {\it self-tuned} oscillation (STO) has been reported by~\citep[]{maity:EPL_2013} in rotating RBC ($\mathrm{Q} = 0$) system at the onset of convection. Here, it is observed that STO persists at the onset for very small values of $\mathrm{Q}$ and it becomes asymmetric ($max|W_{101}| \neq max|W_{011}|$), attributed to the presence external uniform magnetic field in the horizontal direction which breaks the $x \leftrightharpoons y$ symmetry of the system. Figures~\ref{fig:dns_pattern} (a)-(h) display the corresponding pattern dynamics. The amplitude of a set of straight rolls grows in time. But before it reaches to its maximum value a new set of straight rolls, perpendicular to the previous set of rolls is excited. Shortly after that the old set of rolls disappears and does not grow for a finite time. When the amplitude of the new set of rolls gets closer to its maximum value, once again the old set of rolls is excited. These solutions are also called oscillatory cross rolls of type I (OCR-I) in the literature. 

With the increase of $\mathrm{Q}$, STO nature of the solution at the onset is suppressed and oscillating cross rolls are observed.
Note that for this oscillatory cross-rolls $max|W_{011}| > max|W_{101}|$ (figure~\ref{fig:tau_20_diff_Q_onset}(b)). We call it OCR-II$'$ and corresponding pattern dynamics is shown in figures~\ref{fig:dns_pattern} (i)-(k). Stationary cross-rolls (CR$'$) i.e. two sets of mutual perpendicular rolls appear at the onset for $\mathrm{Q} = 1.1$ (figure~\ref{fig:tau_20_diff_Q_onset}(c)), for which $|W_{011}| > |W_{101}|$. Pattern for CR$'$ is shown in figure~\ref{fig:dns_pattern}(l). As $\mathrm{Q}$ is increased further we again observe oscillatory cross-rolls at the onset for $\mathrm{Q} = 10$. But this time $max|W_{101}| > max|W_{011}|$ (figure~\ref{fig:tau_20_diff_Q_onset}(d)) and we refer it as OCR-II. Pattern dynamics corresponding to OCR-II are shown in figures~\ref{fig:dns_pattern} (m)-(o). For $\mathrm{Q} = 16$, stationary cross-rolls (CR) appears at the onset for which $|W_{101}| > |W_{011}|$(figure~\ref{fig:tau_20_diff_Q_onset}(e)). Figure~\ref{fig:dns_pattern} (p) shows the convective pattern corresponding to CR solution. Further increasing the value of $\mathrm{Q}$ we observe steady 2D rolls ($W_{101} \neq 0, W_{011} = 0$) at the onset of convection (figure~\ref{fig:tau_20_diff_Q_onset}(f)).      
\begin{figure}
\begin{center}
\includegraphics[height=!, width=\textwidth]{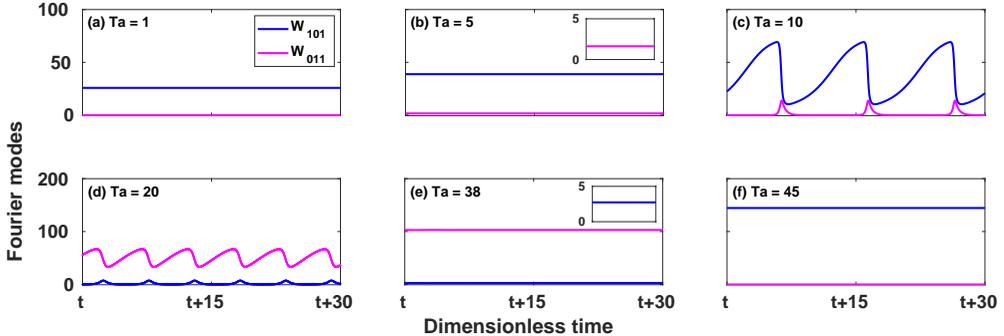}
\end{center}
\caption{Different flow regimes are observed at the onset of convection ($r = 1.001$) for $\mathrm{Q} = 1$ and different values of $\mathrm{Ta}$. Figure displays the temporal evolution of two largest Fourier modes $W_{101}$ and $W_{011}$ corresponding to the different solutions.}
\label{fig:Q_1_tau_diff_onset}
\end{figure}

Now to understand the effect of rotation on the onset of convection in presence of weak external uniform horizontal magnetic field,  we fix $\mathrm{Q} = 1$ and increase $\mathrm{Ta}$ starting from a small value. 
%It is well known that external magnetic field in horizontal direction does not affect the onset of convection and it is unchanged from the non-magnetic case. Therefore steady 2D rolls is observed at the onset of convection for low Prandtl-number fluids. Varying the value of $\mathrm{Ta}$ we get various flow patterns at the onset of convection. 
Temporal variation of two largest Fourier modes $W_{101}$ and $W_{011}$ corresponding to different flow patterns observed in DNS  are shown in figure~\ref{fig:Q_1_tau_diff_onset}. For very small value of $\mathrm{Ta}$ i.e. $\mathrm{Ta} = 1$ we observe straight 2D rolls ($W_{101} \neq 0, W_{011} = 0$) at the onset (see fig~\ref{fig:Q_1_tau_diff_onset} (a)). As we increase $\mathrm{Ta}$, we  observe stationary cross-rolls (CR) (fig~\ref{fig:Q_1_tau_diff_onset} (b)), oscillatory cross-rolls i.e. OCR-II and OCR-II$'$  (fig~\ref{fig:Q_1_tau_diff_onset} (c) and (d)), stationary cross-rolls (CR$'$) (fig~\ref{fig:Q_1_tau_diff_onset} (e)) and steady 2D rolls ($W_{101} \neq 0, W_{011} = 0$) (fig~\ref{fig:Q_1_tau_diff_onset} (f)) at the onset of convection. 

Form the DNS results discussed above, we note that as $\mathrm{Ta}$ and $\mathrm{Q}$ are varied in a small range, various interesting stationary and time-dependent flow patterns appear near the onset of convection. We try to understand the origin of these different solutions just at the onset of convection by constructing a low-dimensional model from the DNS data and performing bifurcation analysis of the model. The details of the low dimensional modeling and subsequent analysis have been discussed below.

\subsection{A low dimensional model}
To derive a low-dimensional model for the above mentioned purpose we follow the procedure described in~\citep{nandu:2016}. The main idea behind the procedure is to identify the large scale modes by calculating contributions of an individual mode to the total energy from the DNS data. From the energy computation of the modes, here we identify eleven large-scale vertical velocity modes: $W_{101}$, $W_{011}$, $W_{112}$, $W_{211}$, $W_{121}$, $W_{202}$, $W_{022}$, $W_{301}$, $W_{031}$, $W_{103}$, $W_{013}$, sixteen vertical vorticity modes: $Z_{101}$, $Z_{011}$, $Z_{110}$, $Z_{112}$, $Z_{211}$, $Z_{121}$, $Z_{310}$, $Z_{130}$, $Z_{202}$, $Z_{022}$, $Z_{103}$, $Z_{013}$, $Z_{301}$, $Z_{031}$, $Z_{020}$, $Z_{200}$ and twelve temperature modes: $T_{101}$, $T_{011}$, $T_{112}$, $T_{211}$, $T_{121}$, $T_{202}$, $T_{022}$, $T_{301}$, $T_{031}$, $T_{103}$, $T_{013}$, $T_{002}$. Hence the expression for $v_3$, $\omega_3$ and $\theta$ becomes:
\begin{eqnarray}
v_3 &=& [W_{101}(t)\cos{k_c x} + W_{011}(t)\cos{k_c y}]\sin{\pi z} + W_{112}(t)\cos{k_c x}\cos{k_c y}\sin{2\pi z}\nonumber\\
&+& [W_{202}(t)\cos{2k_c x} + W_{022}(t)\cos{2k_c y}]\sin{2\pi z} + W_{121}(t)\cos{k_c x}\cos{2k_c y}\sin{\pi z}\nonumber\\
&+& [W_{301}(t)\cos{3k_c x} + W_{031}(t)\cos{3k_c y}]\sin{\pi z} + W_{211}(t)\cos{2k_c x}\cos{k_c y}\sin{\pi z}\nonumber\\
&+& [W_{103}(t)\cos{k_c x} + W_{013}(t)\cos{k_c y}]\sin{3\pi z},\\
\nonumber\\
\omega_3 &=& [Z_{101}(t)\cos{k_c x} + Z_{011}(t)\cos{k_c y}]\cos{\pi z} + Z_{310}(t)\sin{3k_c x}\sin{k_c y}\nonumber\\
&+&[Z_{202}(t)\cos{2k_c x} + Z_{022}(t)\cos{2k_c y}]\cos{2\pi z} + Z_{110}(t)\sin{k_c x}\sin{k_c y}\nonumber\\ 
&+& [Z_{301}(t)\cos{3k_c x} + Z_{031}(t)\cos{3k_c y}]\cos{\pi z} + Z_{112}(t)\sin{k_c x}\sin{k_c y}\cos{2\pi z}\nonumber\\
&+& [Z_{103}(t)\cos{k_c x} + Z_{013}(t)\cos{k_c y}]\cos{3\pi z} + Z_{211}(t)\sin{2k_c x}\sin{k_c y}\cos{\pi z}\nonumber\\
&+& Z_{020}(t)\cos{2k_c y}+ Z_{200}(t)\cos{2k_c x} + Z_{121}(t)\sin{k_c x}\sin{2k_c y}\cos{\pi z}\nonumber\\
&+& Z_{130}(t)\sin{k_c x}\sin{3k_c y},\\
\nonumber\\
\theta &=& [T_{101}(t)\cos{k_c x} + T_{011}(t)\cos{k_c y}]\sin{\pi z} + T_{112}(t)\cos{k_c x}\cos{k_c y}\sin{2\pi z}\nonumber\\
&+& [T_{202}(t)\cos{2k_c x} + T_{022}(t)\cos{2k_c y}]\sin{2\pi z} + T_{121}(t)\cos{k_c x}\cos{2k_c y}\sin{\pi z}\nonumber\\
&+& [T_{301}(t)\cos{3k_c x} + T_{031}(t)\cos{3k_c y}]\sin{\pi z} + T_{211}(t)\cos{2k_c x}\cos{k_c y}\sin{\pi z}\nonumber\\
&+& [T_{103}(t)\cos{k_c x} + T_{013}(t)\cos{k_c y}]\sin{3\pi z} + T_{002}\sin{2\pi z}.
\end{eqnarray}
Horizontal components of the velocity $v_1$ and $v_2$ are then determined using the equation of continuity. Projecting the hydrodynamic system~(\ref{eq:heat})-~(\ref{eq:magnetic1}) on these modes, we get $39$ coupled nonlinear ordinary differential equations which is the low dimensional model for the investigation of the bifurcation structures near the onset of convection. We integrate the model using standard $RK4$ integration scheme. The model shows good qualitative agreement with DNS in our considered ranges of $\mathrm{Ta}$ and $\mathrm{Q}$. In the next section, we discuss these in detail.

\subsection{Analysis of the model and DNS results}
\begin{figure}
\begin{center}
\includegraphics[height=!, width=\textwidth]{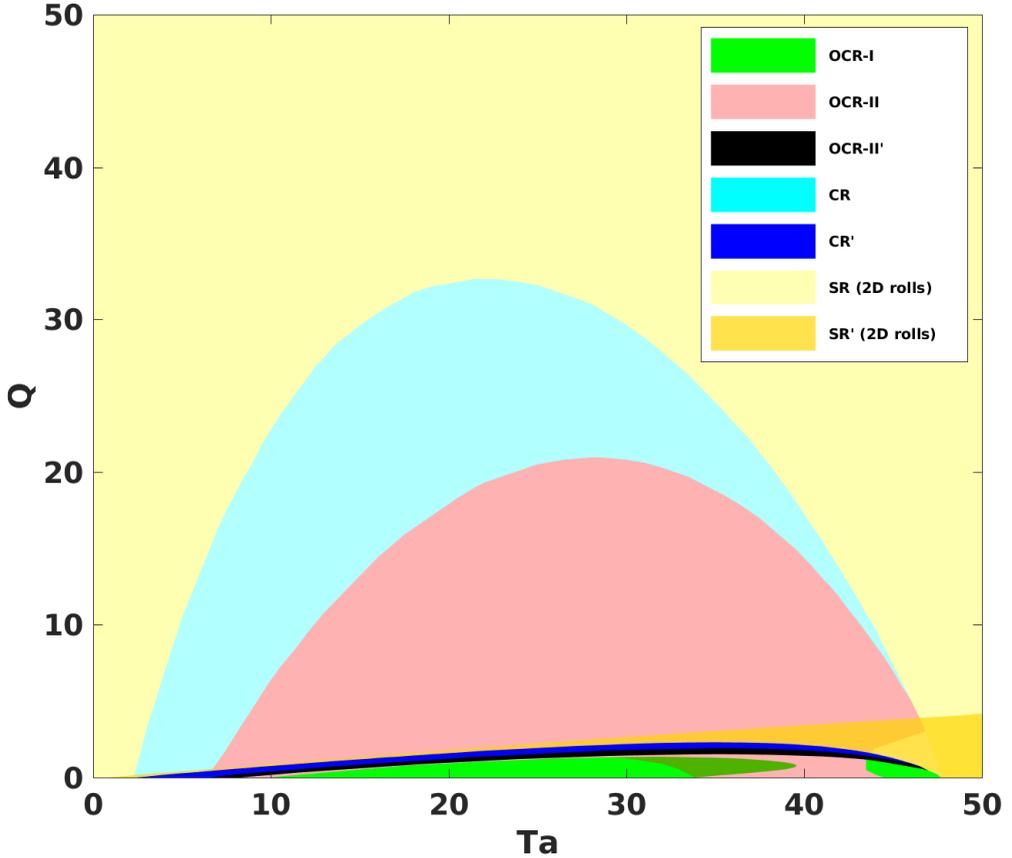}
\end{center}
\caption{Two-parameter diagram constructed from model to demonstrate convective flow patterns near primary instability ($r = 1.001$) for $\mathrm{Pr} = 0.025$ depending on the different values of $\mathrm{Ta}$ and $\mathrm{Q}$ in the range $\mathrm{Ta} \leq 50$, $\mathrm{Q} \leq 50$. Different colors have been used to represent different solutions. The faded and deep yellow regions show steady 2D rolls solutions, in the faded region we have, $W_{101} \neq 0$ and $W_{011} = 0$, whereas $W_{101} = 0$ and $W_{011} \neq 0$ in the deep region. CR and CR$'$ solutions are denoted by cyan and blue colored regions respectively. The faded red region represents OCR-II solutions and the black region below the blue region shows OCR-II$'$ solutions. The green islands near bottom display the OCR-I solutions.}
\label{fig:pr0p025_two_par_tau_Q}
\end{figure}

To probe the origin of different flow regimes observed in DNS very close to the onset of convection, we first construct a two parameter diagram (see figure~\ref{fig:pr0p025_two_par_tau_Q}) for $\mathrm{Pr} = 0.025$ using the $39$-mode model near the onset of convection ($r = 1.001$) in $\mathrm{Ta}-\mathrm{Q}$ plane. Both $\mathrm{Ta}$ and $\mathrm{Q}$ are varied in the range $0 - 50$. 
Depending on the values of the parameters in the chosen range and choice of initial conditions, different flow regimes are obtained which are shown with different colors in the figure. 

The green islands near the bottom of the figure represent OCR-I solutions, which start near $\mathrm{Ta} = 10$ and as $\mathrm{Ta}$ is varied, they exist in the ranges $10\leq \mathrm{Ta} \leq 39.6$ and $43.5 \leq \mathrm{Ta} \leq 47.6$ for small values of $\mathrm{Q}$. The black region above the green region which starts near $\mathrm{Ta} = 7$ and ends at $\mathrm{Ta} = 47$ displays OCR-II$'$ solutions. CR$'$ solutions are shown with blue color in the figure. This flow regime exists in the range $2.5 \leq \mathrm{Ta} \leq 47$. From the figure it is apparent that both OCR-II$'$ and CR$'$ regimes exist at the onset for small values of $\mathrm{Q}$. 
%Width of these two regimes increase as we raise the value of $\mathrm{Ta}$ and become most near $\mathrm{Ta} = 35$. At $\mathrm{Ta} = 35$, OCR-II$'$ solutions exist in the range of $1.52 \leq \mathrm{Q} \leq 1.92$ and CR$'$ solutions exist in the range $1.92 \leq \mathrm{Q} \leq 2.03$. Width of both regions decrease gradually for subsequent higher values of $\mathrm{Ta}$ and both regions vanish after $\mathrm{Ta} = 47$. 
The region of existence of 2D rolls solutions along $x$-axis (SR$'$, the straight rolls) has been marked with deep yellow color in the figure~\ref{fig:pr0p025_two_par_tau_Q}. This flow regime starts at $\mathrm{Ta} = 1$ and continues for all subsequent increased values of $\mathrm{Ta}$ for small values of $\mathrm{Q}$. 
Width of this flow regime also increases with $\mathrm{Ta}$. The faded red region represents OCR-II solutions which exists in the range $6.5 \leq \mathrm{Ta} \leq 46.8$. The upper boundary of the region increases as we increase the value of $\mathrm{Ta}$ till $\mathrm{Ta} = 28$ and then decreases as the value of $\mathrm{Ta}$ is increased further. The lower boundary lies on the upper boundary of OCR-I solutions regime till $\mathrm{Ta} = 29$, then differs from it and touches the horizontal axis at $\mathrm{Ta} = 34$. Again it starts to grow at $\mathrm{Ta} = 44$ and the whole region ends at $\mathrm{Ta} = 46.8$. From the figure, we see that faded red region overlaps with some portions of green, black, blue and deep yellow regions. As a result, we get multiple solutions on those regions depending on the choice of initial conditions. The cyan region shows CR solutions and exists in the range $2.3 \leq \mathrm{Ta} \leq 46$. Width of this flow regime also grows as the value of $\mathrm{Ta}$ is raised to $\mathrm{Ta} = 22$ and then decreases for the subsequent higher values of $\mathrm{Ta}$. Finally, the faded yellow region displays the steady 2D rolls solutions along $y$-axis (SR, the straight rolls). This flow regime exists for the higher values of $\mathrm{Q}$ as we increase the value of $\mathrm{Ta}$ till $\mathrm{Ta} = 22$ and then as the lower boundary of the region decreases for higher values of $\mathrm{Ta}$, the regime continues to exist for lower values of $\mathrm{Q}$. Near $\mathrm{Ta} = 48$ the lower boundary of the region touches the horizontal axis and the region continues to exist from $\mathrm{Q} = 0$. An interesting scenario occurs in the range $48 \leq \mathrm{Ta} \leq 50$ and $\mathrm{Q} \leq 4$ where two yellow regions coexist. Therefore, we get two different 2D rolls solutions in that region depending on the choice of initial conditions. DNS also exhibits qualitatively similar results in the considered parameter range.

Now to understand the origin of different flow regimes at the onset of convection, we construct bifurcation diagrams using the low-dimensional model by choosing specific values of $\mathrm{Ta}$ and $\mathrm{Q}$ from qualitatively different flow regimes as observed in the figure~\ref{fig:pr0p025_two_par_tau_Q}. 
%We choose $\mathrm{Ta} = 20$ so that we can explore the origin of different solutions by varying the value of $\mathrm{Q}$. To discover the origin of remaining solutions we vary the value of $\mathrm{Ta}$ for $\mathrm{Q} = 1$. 

\subsubsection{Bifurcation structures for weak magnetic field ($\mathrm{Q} < 5$)}
\begin{figure}
\begin{center}
\includegraphics[height=!, width=\textwidth]{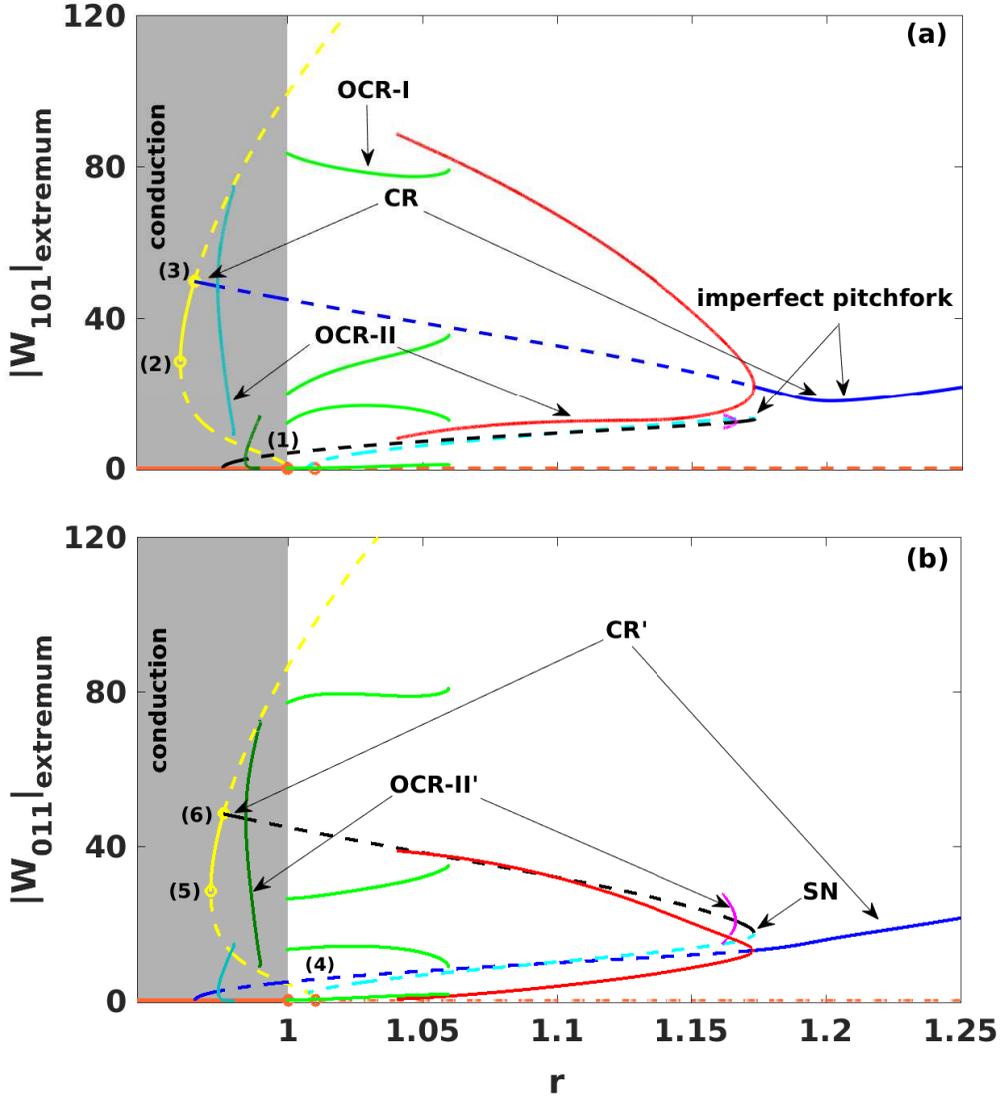}
\end{center}
\caption{Bifurcation diagrams constructed using the model for $\mathrm{Ta} = 20$ for $\mathrm{Q} = 0.5$ and for $\mathrm{Pr} = 0.025$. Extremum values of two variables $|W_{101}|$ and $|W_{011}|$ for different solutions have been shown as a function of $r$ in the range $0.95 \leq r \leq 1.25$ with different curves and different colors in two separate figures. Figure (a) shows bifurcation diagram for the specified parameter values where extremum values of $|W_{101}|$ have been plotted as a function of $r$, whereas extremum values of $|W_{011}|$ have been plotted in figure (b) as a function of $r$. Solid curves represent the stable solutions and dashed curves represent the unstable solutions. Orange curves in both figures represent the conduction state. The yellow curve, in figure (a), shows steady 2D rolls along the y-axis and in figure (b) shows steady 2D rolls along the x-axis. Blue and black curves in both figures display cross rolls dominant along the y-axis (CR) and cross rolls dominant along the x-axis (CR$'$). The cyan curves in both figures show the reminiscence of the stationary square saddle (SQR). Red and pink curves in both figures display oscillatory cross rolls solutions OCR-II and OCR-II$'$ respectively. Grey shaded regions in both figures are the conduction region. The sea-green curves and deep green curves in both figures are also oscillatory cross rolls solutions OCR-II and OCR-II$'$ respectively. Green curves in both figures are the OCR-I type solutions. Orange filled circles in both figures represent subcritical pitchfork bifurcation points at $r = 1$. Empty and filled yellow circles in both figures are the saddle-node points and branch points on 2D rolls, whereas empty orange circles in both figures are the branch points on unstable conduction solutions. Saddle-node (SN) and imperfect pitchfork bifurcations are also shown.}
\label{fig:tau20_Q0p5_bif}
\end{figure}

We start the bifurcation analysis with $\mathrm{Ta} = 20$ and $\mathrm{Q} = 0.5$, for which OCR-I solution appears at the onset of convection. In order to perform detailed bifurcation analysis we construct two separate bifurcation diagrams, where extremum values of $|W_{101}|$ (fig~\ref{fig:tau20_Q0p5_bif}(a)) and $|W_{011}|$ (fig~\ref{fig:tau20_Q0p5_bif}(b)) have been shown for different solutions as a function of $r$, along with their stability informations in the range $0.95 \leq r \leq 1.25$. The conduction state in both diagrams is shown with the orange curve which is stable (solid curve) for $r \leq 1$. It becomes unstable (dashed curve) at $r = 1$ through a subcritical pitchfork bifurcation and an unstable 2D roll branch is originated there for which $W_{101} \neq 0$, $W_{011} = 0$. The subcritical pitchfork bifurcation point is represented by an orange filled circle in both figures and labeled with $(1)$ in figure (a). The yellow curve in figure (a) displays steady 2D rolls along the $y$-axis, which is unstable (dashed) initially and moves backward. This branch, passes through a saddle-node bifurcation point at $r = 0.9604$, becomes stable (solid) after that and starts to move forward from there. The yellow empty circle in figure (a) on 2D roll branch, labeled with $(2)$ shows the saddle-node bifurcation point. Later, a stable cross roll (CR) branch generates from a branch point (filled yellow circle) of the 2D roll branch. The 2D roll branch becomes unstable (dashed) then and remains unstable after that. However, the unstable 2D roll branch comes out of conduction region (shaded grey region) at $r = 1$. Note that $W_{011} = 0, W_{101}\neq 0$ for this 2D rolls solution branch. As a result, this branch is absent in figure (b). Another 2D roll branch for which $W_{101} = 0$, $W_{011} \neq 0$, is originated from a branch point located on the unstable conduction solution branch at $r = 1.0103$. This branch point is shown with an empty orange circle in both figures and labeled with $(4)$ in figure (b). The 2D roll branch is shown with yellow curve in figure (b) and is not present in figure (a), since $W_{101} = 0$ for this branch. This 2D roll branch is also initially unstable, moves backward and exists for reduced values of $r$. Like the previous one, this 2D roll branch also goes through a saddle-node bifurcation at $r = 0.9715$. The bifurcation point is displayed by an empty yellow circle on the 2D roll branch in figure (b) and labeled with (5). After that, the branch becomes stable and continues to move forward. At $r = 0.9762$, a stable cross roll (CR$'$) branch is generated from a branch point of the 2D roll branch and the 2D roll branch becomes unstable thereafter. A yellow filled circle is used to denote this branch point and labeled as $(6)$ in figure (b).  
This unstable 2D roll branch leaves the conduction region as soon as the value of $r$ reaches $r = 1$ and continues to exist for higher values of $r$. The blue curves in both figures represent the CR solutions which originate from a branch point of 2D rolls solution along $y$-axis. Note that for CR solutions we have $W_{101} \neq 0$, $W_{011} \neq 0$ and $|W_{101}| \geq |W_{011}|$. The stable CR branch undergoes through a Hopf bifurcation at $r = 0.9742$, becomes unstable and stable limit cycle (sea-green curve) appears. However the limit cycle exists in the conduction region only. The unstable CR branch then becomes stable via an inverse Hopf bifurcation at $r = 1.1731$ and continues exist for higher values of $r$ till $r = 1.25$. The stable CR$'$ branch (black curve) for which $W_{101} \neq 0$, $W_{011} \neq 0$ and $|W_{101}| \leq |W_{011}|$, also becomes unstable via a Hopf bifurcation at $r = 0.9844$. As a result, stable limit cycle (deep green curve) is generated there. However, this limit cycle remains in the conduction region only. The unstable CR$'$ branch then becomes stable again via an inverse Hopf bifurcation at $r = 1.1667$. This stable CR$'$ branch shows a saddle-node bifurcation at $r = 1.1735$ as the value of $r$ is increased and meets an unstable CR$'$ branch (cyan dashed curve). Figure (b) demonstrates a clear view of the bifurcation scenario of CR$'$ branch. 

Now if we look at both figures from the right-hand side a scenario of imperfect pitchfork bifurcation appears in both figures near $r = 1.194$. The imperfection is due to the presence of an external uniform horizontal magnetic field which breaks the $x \leftrightharpoons y$ symmetry of the system. As a result, we have two different 2D rolls solutions and two different cross rolls solutions which are not symmetrically connected. In the absence of magnetic field ($\mathrm{Q} = 0$), these two rolls would coincide and stable CR and CR$'$ branches would meet at a supercritical pitchfork bifurcation point. Comparing the magnetic and non-magnetic cases we realize that cyan curves are the reminiscence of the square saddle which would exist when $\mathrm{Q} = 0$ and we denote it by SQR (see~\citep[]{maity:EPL_2013})

As we pointed out earlier CR and CR$'$ branches pass through supercritical Hopf bifurcation points at $r = 1.1731$ and $r = 1.1667$ respectively and stable limit cycles are generated there. The limit cycles generated via the Hopf bifurcation of the CR and CR$'$ branches are the OCR-II (solid red curves) and OCR-II$'$ (solid pink curves) solutions respectively, as they show oscillatory cross roll patterns dominated along the $y$-axis and along the $x$-axis. Pattern dynamics corresponding to these solutions are similar to what we observe in DNS. Note that the CR and CR$'$ branches undergo supercritical Hopf bifurcations in the conduction region also. Therefore, the sea-green curves and deep-green curves in both figures are also OCR-II and OCR-II$'$ solutions respectively. However as mentioned earlier these OCR-II and OCR-II$'$ solutions remain in the conduction region only and very difficult to track them from DNS. One thing we would like to mention here that the OCR-II and OCR-II$'$ solutions are not connected by any symmetry of the system, instead they exist independently. As the value of $r$ is decreased, the OCR-II$'$ limit cycle, represented by pink curves, becomes homoclinic to the SQR saddle at $r = 1.1619$ and ceased to exist. While the OCR-II limit cycle, displayed with red curves, continues to exist and becomes homoclinic to the SQR saddle at $r = 1.0405$. Immediately after this homoclinic bifurcation, as the value of $r$ is reduced, asymmetric glued limit cycles (green curves) appear and continue to exist till $r = 1$. This is a scenario of the imperfect gluing bifurcation, which we will discuss in a separate subsection.

%\subsection{Bifurcation structure for $\mathrm{Q} = 2$}
Next we perform the bifurcation analysis for $\mathrm{Ta} = 20$ and $\mathrm{Q} = 2$ for which we have OCR-II solution at the onset. In this case, also we prepare two separate bifurcation diagrams in order to perform bifurcation analysis properly. From bifurcation diagrams, we observe that a small change in the strength of the magnetic field brings significant changes in the bifurcation structure of the system. 
\begin{figure}
\begin{center}
\includegraphics[height=!, width=\textwidth]{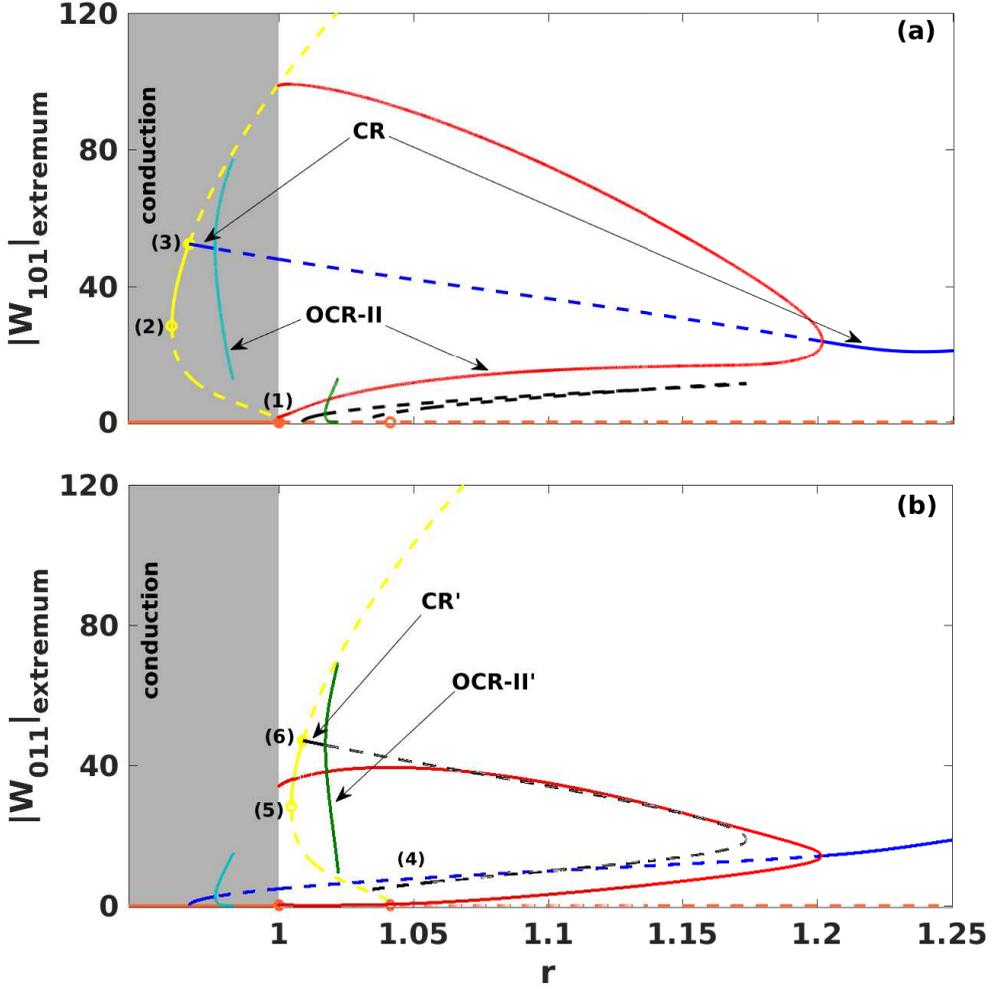}
\end{center}
\caption{Bifurcation diagrams computed for $\mathrm{Ta} = 20$ for $\mathrm{Q} = 2$ and for $\mathrm{Pr} = 0.025$ using the model in the range $0.95 \leq r \leq 1.25$. Stable and unstable solutions in both figures are denoted by solid and dashed curves. Shaded grey regions in both figures are the conduction region. Conduction solutions in each figure are represented by orange curves. Filled and empty orange circles on it are the subcritical pitchfork bifurcation point and branch point respectively. Yellow curves in both figures represent steady 2D rolls: in (a) along the y-axis and in (b) along the x-axis. Empty and filled yellow circles on these 2D rolls are the saddle-node bifurcation point and branch point respectively. Blue and black curves in both figures show the CR and CR$'$ solutions. The OCR-II solutions in both figures are denoted by sea-green and red colored curves, whereas deep green curves in both figures represent the OCR-II$'$ solutions. }
\label{fig:tau20_Q2_bif}
\end{figure}

The steady 2D roll branch along $y$-axis, CR branch and two different OCR-II solution branches behave similar as they have been for $\mathrm{Q} = 0.5$, except the OCR-II solution branch, represented by red curves, never becomes homoclinic and continues to exist till $r = 1$ (see figure~\ref{fig:tau20_Q2_bif}(a)). It is clear from figure~\ref{fig:tau20_Q2_bif}(b) that there is a significant change in the bifurcation structure of 2D roll solution branch along $x$-axis and CR$'$ solution branch. The yellow curve in figure~\ref{fig:tau20_Q2_bif}(b) represents the 2D rolls solution along $x$-axis shows similar behavior like it has been for $\mathrm{Q} = 0.5$, i.e., initially unstable, moves backward, turns around and becomes stable via a saddle-node bifurcation and becomes unstable from a branch point on it, from which a stable CR$'$ branch emerges. But this time the branch originates from a branch point (orange empty circle) of the unstable conduction solution branch which is far away from the onset of convection compare to the case of $\mathrm{Q} = 0.5$ and remains out of conduction region. Also the width of the stable portion of the 2D roll branch i.e., the distance between the points marked as $(5)$ and $(6)$ in figure~\ref{fig:tau20_Q2_bif}(b) becomes smaller than the case of $\mathrm{Q} = 0.5$. Now the stable CR$'$ branch goes through a Hopf bifurcation and stable limit cycle appears, which is the OCR-II$'$ solution (deep-green curves). The stable CR$'$ branch becomes unstable after the Hopf bifurcation and remains unstable thereafter. The OCR-II$'$ limit cycle becomes homoclinic to unstable 2D rolls solution at $r = 1.0219$ and ceased to exist. However, we do not have any gluing bifurcation for this set of parameter values. From figure~\ref{fig:tau20_Q2_bif}(b), we see that in the range $1.0045 \leq r \leq 1.0219$ OCR-II solutions coexist with the 2D rolls solutions along x-axis, CR$'$ solutions and OCR-II$'$ solutions. But the 2D rolls solutions along the x-axis and its associated different solutions i.e., CR$'$ and OCR-II$'$ solutions exist for a different set of initial conditions. DNS shows qualitatively similar behavior as we vary the value of $r$ in the range $0.95 \leq r \leq 1.25$ for the specified set of parameter values.

\subsubsection{Bifurcation structure for higher values of $\mathrm{Q}$}
As we gradually increase, the strength of the magnetic field i.e., the value of $\mathrm{Q}$, substantial changes occur in the bifurcation scenario of the system. To illustrate these changes we construct two more bifurcation diagrams for $\mathrm{Q} = 5$ and $\mathrm{Q} = 50$, for fixed $\mathrm{Ta} = 20$, which are qualitatively different from each other. First we discuss the bifurcation diagram for $\mathrm{Ta} = 20$ and $\mathrm{Q} = 5$ (see figure~\ref{fig:tau20_Q5_bif}). 
\begin{figure}
\begin{center}
\includegraphics[height=!, width=\textwidth]{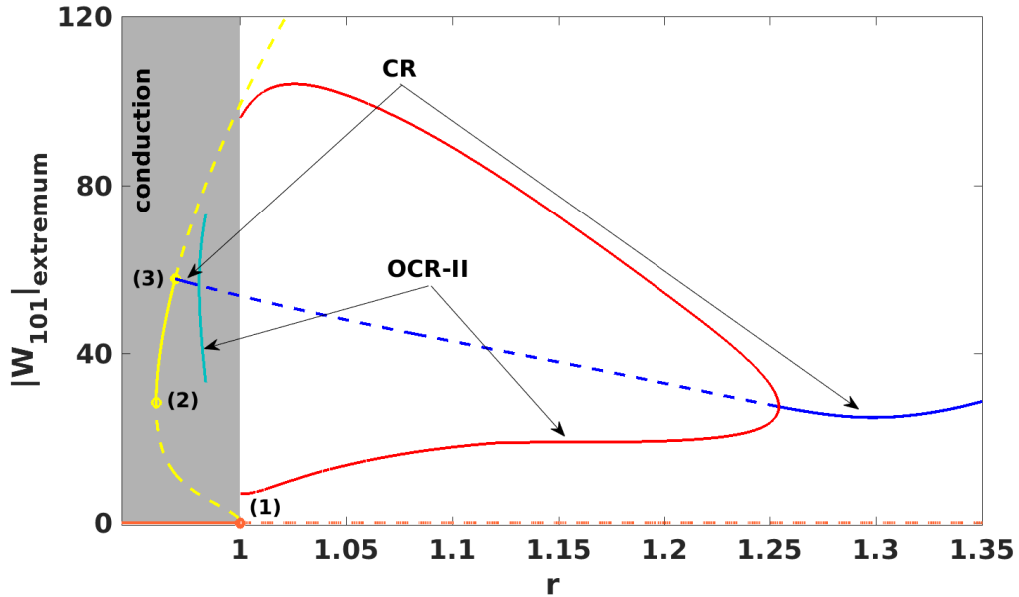}
\end{center}
\caption{Bifurcation diagram created using the model for $\mathrm{Ta} = 20$ for $\mathrm{Q} = 5$ and for $\mathrm{Pr} = 0.025$ in the range $0.95 \leq r \leq 1.35$ by plotting the extremum value of $W_{101}$ as a function of $r$. Same color coding has been used for different solutions as it has been used in the previous diagram.}
\label{fig:tau20_Q5_bif}
\end{figure}
In this bifurcation diagram only the extremum values of $W_{101}$, for different solutions, as a function of $r$ in the range $0.95 \leq r \leq 1.35$ are shown. The reason behind this, as we increase the value of $\mathrm{Q}$, the width of the stable region of the branch of 2D rolls solution along $x$-axis, which is shown in previous two bifurcation diagrams, becomes smaller. In the process, the part of the bifurcation structures associated with this branch and the branch itself vanish for higher values of $\mathrm{Q}$.
In the bifurcation diagram, we notice that the 2D roll branch exhibits similar behavior as it has shown in previous two diagrams. We do not see any movement of the saddle-node bifurcation point of the 2D roll branch (empty yellow circle) as we change the value of $\mathrm{Q}$. However, the branch point located on the 2D roll branch (filled yellow circle), from which CR branch originates, moves away further. As a result, length of the stable portion (distance between $(2)$ and $(3)$) of the 2D roll branch increases. The CR branch also shows the same characteristic behavior as it has shown before and undergoes through two Hopf bifurcations, one inside the conduction region at $r = 0.9807$ and another one for the higher value of $r$ at $r = 1.2541$. Two stable limit cycles appear as a result of these Hopf bifurcations. However, the limit cycle generates from the Hopf point inside the conduction region (sea-green curve) continues to exist in the conduction region only. While the limit cycle, originates from the Hopf point located at higher value of $r$ (red curve) continues to exist for reduced values of $r$ till $r = 1$ and we get OCR-II solution at the onset of convection. 
\begin{figure}
\begin{center}
\includegraphics[height=!, width=\textwidth]{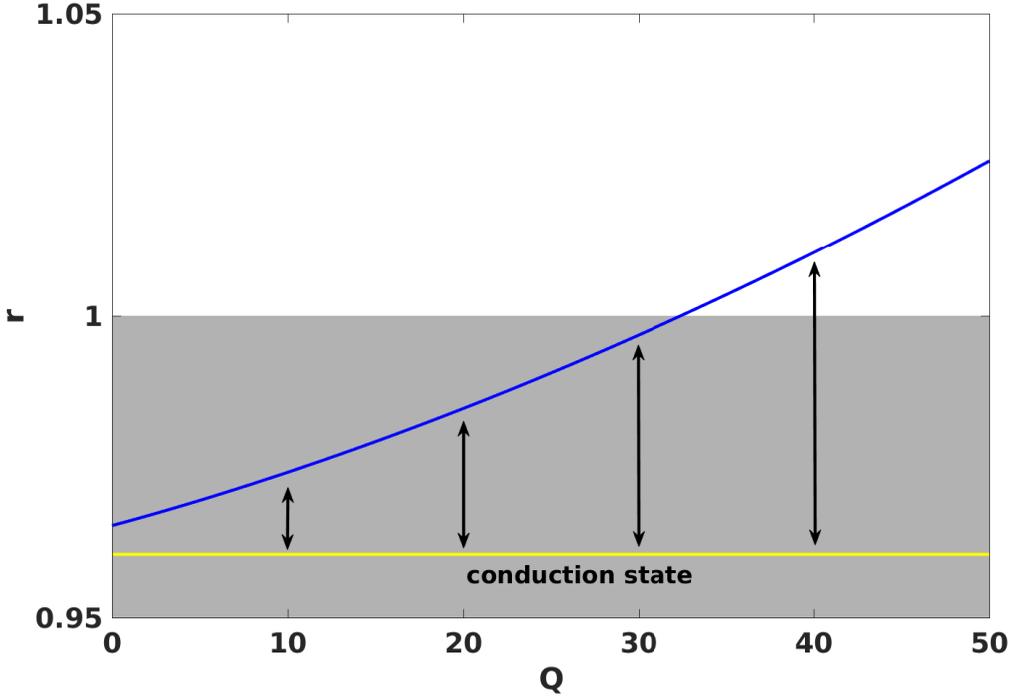}
\end{center}
\caption{Continuation of the saddle-node bifurcation point and the branch point located on 2D roll branch from which CR branch originates is shown in the figure, over the two-parameter space ($\mathrm{Q}$ vs $r$), for fixed $\mathrm{Ta} = 20$. Grey shaded region is the conduction region. The blue curve and the yellow curve represent the continuation of branch point and saddle-node bifurcation point respectively. Arrows are used to illustrate that the distance between them increases as the value of $\mathrm{Q}$ is raised.}
\label{fig:pr0p025_Q_vs_r_twopar}
\end{figure}
As mentioned earlier, the branch point on the 2D roll branch from which CR branch originates keeps moving away from the saddle-node bifurcation point as we increase the value of $\mathrm{Q}$. While the later one remains invariant as the value of $\mathrm{Q}$ is varied for a fixed $\mathrm{Ta}$. As a result, the length of the stable section of the 2D roll branch increases as we raise the value of $\mathrm{Q}$ and stable 2D rolls solution appears at the onset of convection. Figure~\ref{fig:pr0p025_Q_vs_r_twopar} displays the scenario in two-parameter space ($\mathrm{Q}$ vs $r$) for $\mathrm{Ta} = 20$. The continuation of the saddle-node bifurcation point and branch point are shown in this figure. The grey shaded region, as usual, represents the conduction region. The yellow curve is the continuation of the saddle-node bifurcation point in the parameter space. Whereas, the blue curve shows the continuation of the branch point over the parameter space. It is clear from the figure that as  $\mathrm{Q}$ is increased, the distance between the two curves increases which is the length of the stable section of the 2D roll branch. Near $\mathrm{Q} = 32.45$ the blue curve leaves the conduction region and continues to exist there for higher values of $\mathrm{Q}$. As a result, stable 2D rolls solution appears at the onset of convection as soon as the value of $\mathrm{Q}$ crosses $\mathrm{Q} = 32.45$. 
\begin{figure}
\begin{center}
\includegraphics[height=!, width=\textwidth]{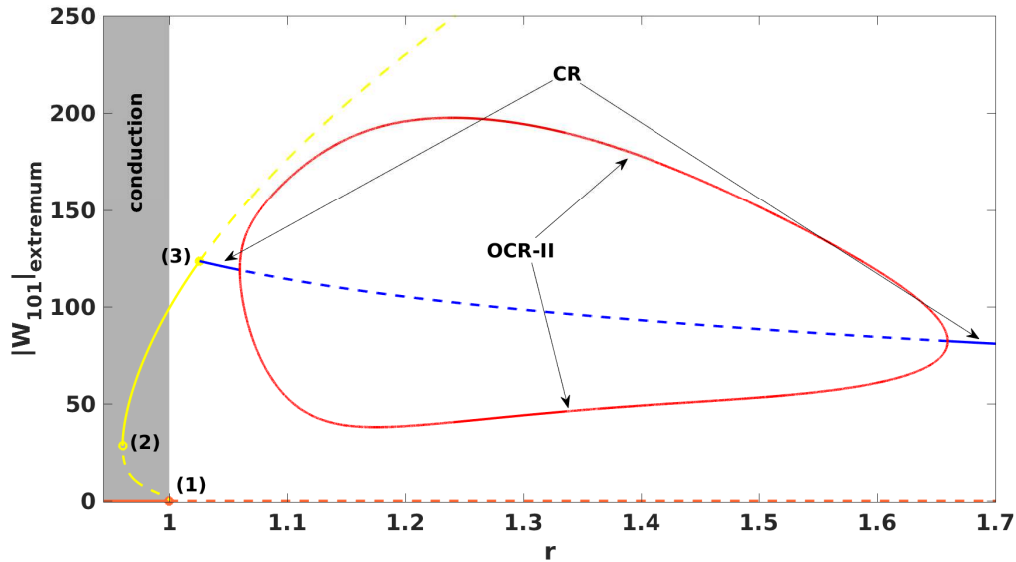}
\end{center}
\caption{Bifurcation diagram obtained from model for $\mathrm{Ta} = 20$ for $\mathrm{Q} = 50$ and for $\mathrm{Pr} = 0.025$. Extremum values of $W_{101}$ is shown as a function of $r$ over the region $0.95 \leq r \leq 1.7$. Solid and dashed curves represent stable and unstable solutions. Orange, yellow, blue and red colored curves show conduction state, steady 2D rolls along the y-axis, CR and OCR-II solution respectively. Conduction region is shown with grey shaded area. Empty yellow circle on the steady 2D rolls is the saddle-node point, whereas filled yellow circle on it and orange filled circle on conduction solution are the branch point and imperfect pitchfork bifurcation point respectively.}
\label{fig:tau20_Q50_bif}
\end{figure}

Now as we turn our attention toward the bifurcation diagram for $\mathrm{Ta} = 20$ and $\mathrm{Q} = 50$ ( see figure~\ref{fig:tau20_Q50_bif}), we see that 2D roll branch comes out of the conduction region and steady 2D rolls solution is observed at the onset of convection. The 2D roll branch then becomes unstable after a branch point appears on it at $r = 1.0258$ from which a CR branch emerges. This CR branch now goes through a supercritical Hopf bifurcation at $r = 1.0597$, becomes unstable and stable limit cycle is generated. This limit cycle grows in size initially and then begins to decrease gradually as we increase the value of $r$. The unstable CR branch then becomes stable via an inverse Hopf bifurcation at $r = 1.66$. An interesting phenomenon, known as ``first order transition" occurs as soon as the stable 2D rolls solution appears at the onset of convection. In the next subsection, we discuss it in detail.

\subsection{First-order transition}
\begin{figure}
\begin{center}
\includegraphics[height=0.4\textwidth, width=\textwidth]{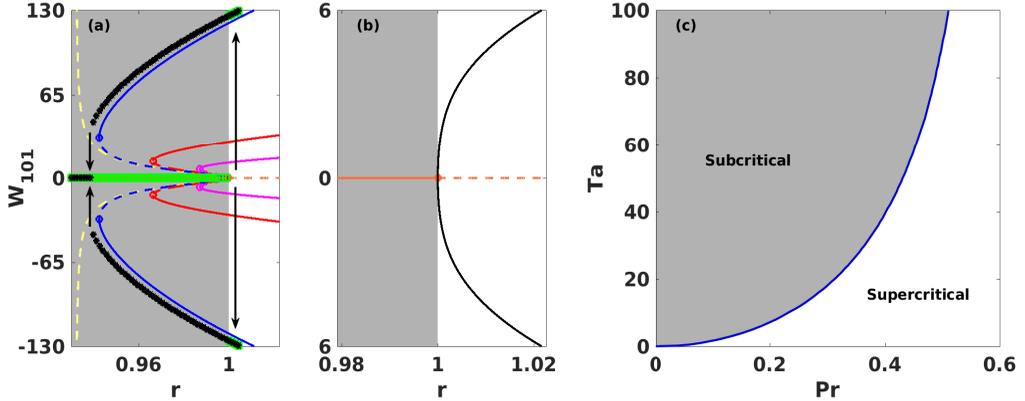}
\end{center}
\caption{Bifurcation diagrams corresponding to first order transition are computed using 3-mode model by varying the parameter $r$ for $\mathrm{Ta} = 20$ shown in figure (a) and (b) for different values of $\mathrm{Pr}$. Solid and dashed lines display the stable and unstable solutions respectively. The shaded grey region is the conduction region. The orange lines in both figure are the conduction solutions which become unstable at $r = 1$. The orange filled circle in figure (a) is the subcritical pitchfork bifurcation point, whereas it represents a supercritical pitchfork bifurcation point in figure (b). Empty circles are the saddle-node bifurcation points for different values of $\mathrm{Pr}$. Yellow, blue, red and pink curves in figure (a) and the black curve in figure (b) are the steady 2D rolls along y-axis for $\mathrm{Pr} = 0$, $\mathrm{Pr} = 0.025$, $\mathrm{Pr} = 0.1$, $\mathrm{Pr} = 0.2$ and $\mathrm{Pr} = 0.4$ respectively. Green filled squares and black filled stars are the results obtained from DNS for $\mathrm{Pr} = 0.025$ and $\mathrm{Ta} = 20$ by moving forward and backward respectively. Figure (c) shows the whether the onset of convection is subcritical or supercritical in $\mathrm{Ta}-\mathrm{Pr}$ plane. The grey shaded region in the figure represents the area where choice of any pair of values of $\mathrm{Ta}$ and $\mathrm{Pr}$, likely to gives subcritical pitchfork bifurcation at the onset of convection whereas choice of any pair of values of $\mathrm{Ta}$ and $\mathrm{Pr}$ in the rest of the diagram gives supercritical pitchfork bifurcation at the onset of the convection.}
\label{fig:diff_tau_Q40_transition}
\end{figure}

First-order transition is a well known phenomenon, which is common to liquid-gas transitions, solid-liquid transitions, superconductors, percolation theory and many other fields~\citep[]{halperin:1974, gunton:1983, binder:1987, kuwahara:1995, parshani:2010, goldenfeld:2018, schrieffer:2018}. Generally it appears together with a hysteresis, that is, a difference between the transition points, as a parameter that drives a transition is increased or decreased. 
\begin{figure}
\begin{center}
\includegraphics[height=0.5\textwidth, width=\textwidth]{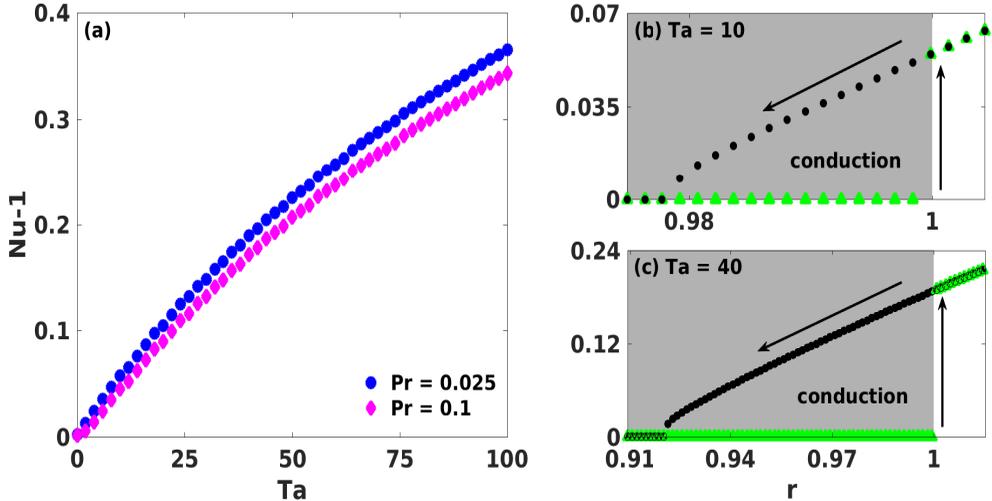}
\end{center}
\caption{Variation of $\mathrm{Nu}-1$ i.e., the ratio of convective heat transport to conductive heat transport at the onset of convection ($r = 1$), computed from DNS, is shown in figure (a) by varying the value of $\mathrm{Ta}$ for $\mathrm{Pr} = 0.025$ and $\mathrm{Pr} = 0.1$ for fixed value of $\mathrm{Q} = 40$. The blue solid circles represent the variation of $\mathrm{Nu}-1$ as a function of $\mathrm{Ta}$ for $\mathrm{Pr} = 0.025$ and the variation of same quantity corresponding to $\mathrm{Pr} = 0.1$ is shown by solid pink diamonds. Figure (b) and (c) display the variation of $\mathrm{Nu}-1$ as a function of $r$ for $\mathrm{Ta} = 10$ and $\mathrm{Ta} = 40$ respectively for $\mathrm{Pr} = 0.025$. The green filled triangles are computed from DNS by moving forward form lower to higher values of $r$. Whereas the black solid circles are plotted by moving backward from higher to lower values $r$ where result of the last simulation is used as the initial condition for current simulation. The shaded grey region in figure (b) and (c) is the conduction region.}
\label{fig:nusselt_number}
\end{figure}
The parameter, which causes the transition is known as order-parameter. Here we observe the first-order transition in our considered system when steady 2D rolls solution appears at the onset of convection. The Prandtl-number plays the role of order-parameter in this case. We start with the results from DNS, where we notice a sudden enhancement in convective heat transport at the onset of convection for a given set of parameter values. Figure~\ref{fig:nusselt_number}(a) illustrates the scenario, where $\mathrm{Nu}-1$ i.e., the ratio of convective heat transport to conductive heat transport at the onset of convection, is plotted by varying $\mathrm{Ta}$ for $\mathrm{Pr} = 0.025$ and $\mathrm{Pr} = 0.1$ for $\mathrm{Q} = 40$. It is clear from the figure that as the value of $\mathrm{Ta}$ is increased for fixed values of $\mathrm{Pr}$ and $\mathrm{Q}$, the jump in the convective heat transport at the onset of convection increases. To find the origin of this abrupt change in convective heat transport at the onset of convection we first perform DNS for higher values of $r (r \geq 1)$ for a given set of values of $\mathrm{Ta}$, $\mathrm{Q}$ and $\mathrm{Pr}$. Then using the final values of all the fields of the last simulation as the current initial conditions we proceed to current simulation by reducing the value of $r$ in a small amount. Following the described method of moving backward, we interestingly observe that convection sustains even for $r < 1$ i.e., below the onset of convection as we decrease the value of $r$, before going to conduction state again. As a result, two separate transition points appear accompanied by a hysteresis loop. Figures~\ref{fig:nusselt_number}(b) and (c) show the first order transition for $\mathrm{Ta} = 10$ and $\mathrm{Ta} = 40$ along with the hysteresis loop for $\mathrm{Pr} = 0.025$ and $\mathrm{Q} = 40$. In both, the figures $\mathrm{Nu}-1$, the convective heat transport, is plotted as a function of $r$. Green filled triangles represent the variation of $\mathrm{Nu}-1$ with $r$ as we move forward (from lower to higher values of $r$), whereas solid black circles display the variation of same quantity with $r$ as we move backward (from higher to lower values of $r$), using the results of the last simulation as the initial conditions for current simulation. In both the figures, we see that the value of $\mathrm{Nu}-1$ is zero for $r < 1$ as we move forward and a jump in the value of $\mathrm{Nu}-1$ appears at $r = 1$ where convection sets in. The scenario becomes different as we move backward starting from $r \geq 1$ for subsequently reduced values of $r$. The quantity $\mathrm{Nu}-1$ remains nonzero and convection persists for $r < 1$. However, $\mathrm{Nu}-1$ decreases gradually as we reduce the value of $r$ in small steps and finally becomes zero at the backward transition point.  The distance between forward and backward transition points is known as the width of the hysteresis loop and by comparing both figures we notice that it depends on the value of $\mathrm{Ta}$ for fixed $\mathrm{Pr}$ and $\mathrm{Q}$. To investigate the bifurcation structure associated with first-order transition we construct a small 3-mode model which gives qualitatively similar results with DNS. We follow the similar procedure to construct the 3-mode model as we have performed before for the 39-mode model. In this case, we neglect all the higher order modes and choose only one velocity mode $W_{101}$, two vorticity modes $Z_{101}$ and $Z_{200}$ and two temperature modes $T_{101}$ and $T_{002}$. Projecting the hydrodynamic system on these modes we get a set consisting of five coupled nonlinear ordinary differential equations. Then we perform the process of adiabatic elimination to eliminate the modes $Z_{200}$ and $T_{002}$. Finally, we are left with a small system containing only three nonlinear ordinary differential equations given by
\begin{eqnarray}
x' &=& -ax - \frac{b}{a}y + \frac{c}{a}z, \nonumber\\
y' &=& bx - ay - \frac{\pi^2}{8(a - \pi^2)} x^{2}y, \nonumber\\
z' &=& \frac{1}{\mathrm{Pr}}x - \frac{a}{\mathrm{Pr}}z - \frac{\mathrm{Pr}}{8}x^{2}z,
\end{eqnarray}
where $x = W_{101}$, $y = Z_{101}$ and $z = T_{101}$. The coefficients are $a = \pi^2 + k_c^2$, $b = \pi\sqrt{\mathrm{Ta}}$ and $c = \mathrm{Ra}k_c^2$, with $k_c$ is the critical wave number for the onset of convection. Note that this model is valid only when steady 2D rolls appears at the onset of convection which typically occurs for sufficiently large values of $\mathrm{Q}$ and for slow rotation rate. When the value of $\mathrm{Q}$ is sufficiently small and time-dependent solutions appear at the onset of convection, the model is unable to capture the associated bifurcation scenario, hence becomes invalid in that case. To uncover the bifurcations associated with first-order transition we create two separate bifurcation diagrams for $\mathrm{Ta} = 20$ corresponding to different values of $\mathrm{Pr}$ (see figures~\ref{fig:diff_tau_Q40_transition} (a) and (b)). Different solutions along with their stability information are shown in both figures for different values of $\mathrm{Pr}$. The conduction state in both figures is shown with an orange curve which is stable for $r < 1$ and loses its stability at $r = 1$. In figure~\ref{fig:diff_tau_Q40_transition} (a), we see that the conduction state becomes unstable through a subcritical pitchfork bifurcation and two symmetrically located unstable 2D roll branches originate from there. These 2D roll branches are shown in figure~\ref{fig:diff_tau_Q40_transition} (a) for different values of $\mathrm{Pr}$ with different colors. The dashed yellow curves represent 2D rolls along $y$-axis corresponding to $\mathrm{Pr} = 0$. This branch starts to move backward from the subcritical pitchfork bifurcation point (solid filled circle in figure~\ref{fig:diff_tau_Q40_transition} (a)) and continues to exist in the conduction region only. As a result, we do not have any first order transition in this case. Now as we increase the value of $\mathrm{Pr}$, a qualitative difference occurs in the bifurcation scenario. The blue curves in figure~\ref{fig:diff_tau_Q40_transition} (a) display the 2D rolls along $y$-axis for $\mathrm{Pr} = 0.025$. These curves initially show behavior like the previous one and start to move backward. But both branches pass through a saddle-node bifurcation at $r = 0.9604$. After that, they become stable, turn around, start to move forward and eventually come out from the conduction region at $r = 1$. Therefore, at the onset of convection, we observe steady 2D rolls with larger amplitude. 

\begin{figure}
\begin{center}
\includegraphics[height=0.55\textwidth, width=\textwidth]{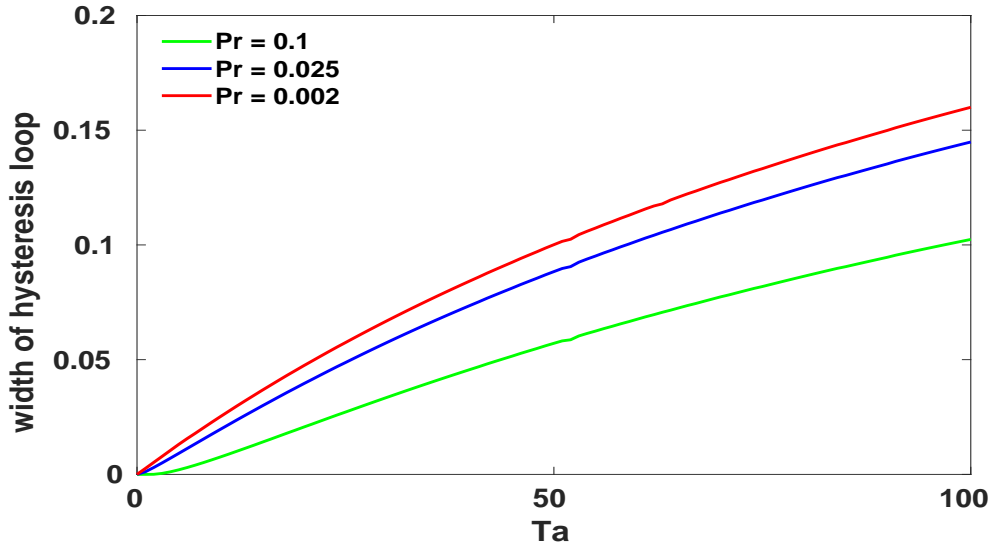}
\end{center}
\caption{The width of the hysteresis loop (distance between forward and backward transition points) is shown as a function of $\mathrm{Ta}$ for different values of $\mathrm{Pr}$ in the figure. Red, blue and green curves represent the width of hysteresis loop corresponding to $\mathrm{Pr} = 0.002$, $\mathrm{Pr} = 0.025$ and $\mathrm{Pr} = 0.1$ respectively in the range $0 \leq \mathrm{Ta} \leq 100$.}
\label{fig:hysteresis_width}
\end{figure}

We notice similar qualitative behavior from DNS also. We get 2D rolls with a large amplitude at the onset of convection when we move forward (green filled squares). After getting the 2D rolls at the onset, if we move backward (black filled stars) using the results of the last simulation as the current initial conditions, we see it follows the stable 2D roll branch which exists in the conduction region until the saddle-node bifurcation point (empty blue circle) before going to zero again. A hysteresis loop along with two different transition points appear as a result of this strange behavior of the 2D roll branch. The subcritical pitchfork bifurcation point, in this case, is the forward transition point and the saddle-node bifurcation point is the backward transition point. Note that the backward transition point is independent of $\mathrm{Q}$ for a fixed value of $\mathrm{Ta}$ and $\mathrm{Pr}$ as we have seen in figure~\ref{fig:pr0p025_Q_vs_r_twopar}. However, from earlier bifurcation analysis we observe that it is $\mathrm{Q}$ which suppresses the time-dependent solutions for small rotation rate and brings the steady 2D rolls at the onset of convection for a fixed value of $\mathrm{Pr}$ so that first-order transition is observed. But once steady 2D rolls appears at the onset of convection, it remains unchanged there as we increase the value of $\mathrm{Q}$ further, for fixed values of $\mathrm{Pr}$ and $\mathrm{Ta}$. The $3$-mode model ensures this phenomenon which does not contain any term related to $\mathrm{Q}$. Now as we increase the value of $\mathrm{Pr}$ gradually for a fixed value of $\mathrm{Ta}$ the width of the hysteresis loop becomes smaller which can be observed in figure~\ref{fig:diff_tau_Q40_transition}(a) where, the red and pink colored curves represent 2D rolls along $y$-axis for $\mathrm{Pr} = 0.1$ and $\mathrm{Pr} = 0.2$ respectively. The variation of the width of hysteresis loop is shown in detail over two parameter space $\mathrm{Ta}$ -- $\mathrm{Pr}$ in figure~\ref{fig:hysteresis_width}. In the figure, the width of the hysteresis loop is displayed as a function of $\mathrm{Ta}$ for three different values of $\mathrm{Pr}$ with three different colors. From the figure, we notice that the width of hysteresis loop increases as we raise the value of $\mathrm{Ta}$ for a fixed value of $\mathrm{Pr}$, and decreases for higher values of $\mathrm{Pr}$ for a fixed value of $\mathrm{Ta}$. Further, increase in the value of $\mathrm{Pr}$ for a fixed value of $\mathrm{Ta}$ brings significant changes at the onset of convection, in which the subcritical pitchfork bifurcation becomes supercritical pitchfork bifurcation and scenario of first-order transition vanishes. Figure~\ref{fig:diff_tau_Q40_transition}(b) shows supercritical pitchfork bifurcation at the onset of convection for $\mathrm{Pr} = 0.4$ for $\mathrm{Ta} = 20$. Transition from subcritical onset to supercritical onset is shown over two parameter space ($\mathrm{Ta}$ -- $\mathrm{Pr}$) in figure~\ref{fig:diff_tau_Q40_transition}(c). The grey shaded region in the figure represents the subcritical onset while the white region represents the supercritical onset. The blue curve in the figure is the marginal curve where the transition from subcritical to supercritical occurs. 

\subsection{Bifurcation structure for higher values of $\mathrm{Ta}$}
\begin{figure}
\begin{center}
\includegraphics[height=!, width=\textwidth]{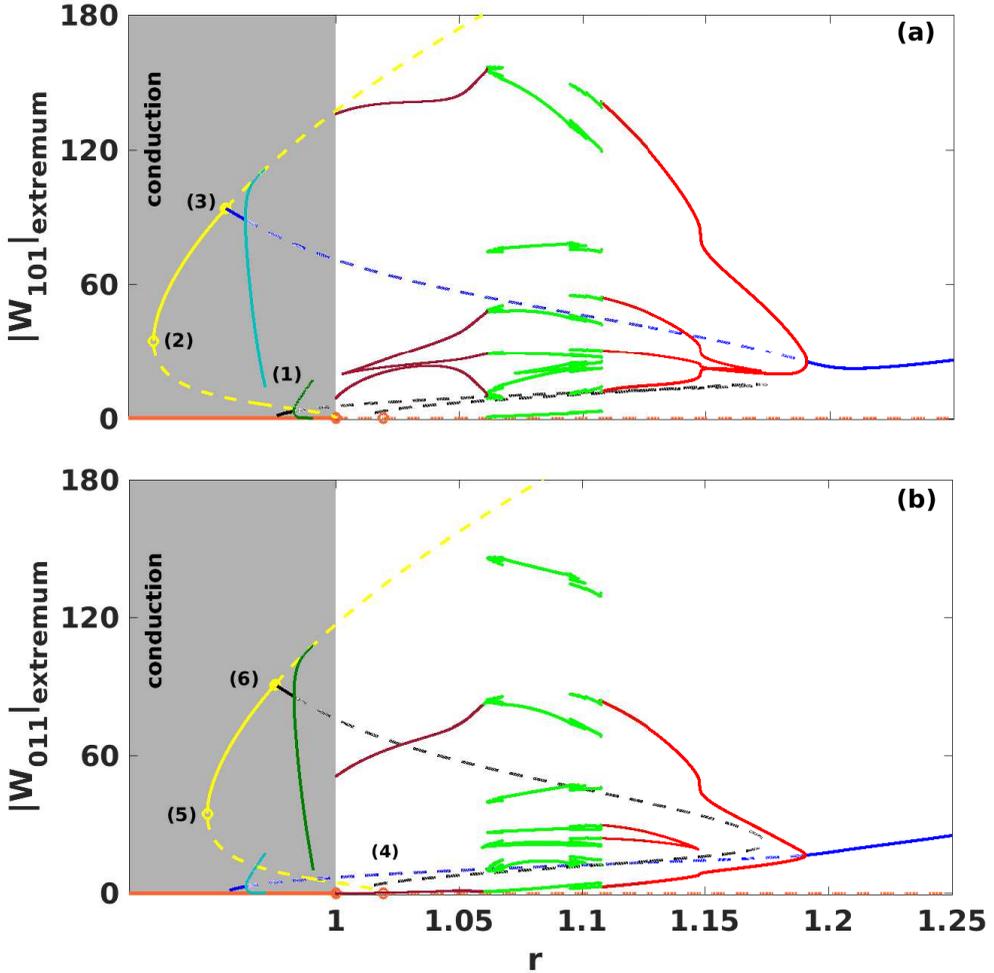}
\end{center}
\caption{Figure shows bifurcation diagram corresponding to $\mathrm{Ta} = 40$, $\mathrm{Q} = 1$ for $\mathrm{Pr} = 0.025$. Same color coding has been used for different solutions as it has been used in the previous bifurcation diagrams except the green dots. The green dots are the extremum of corresponding variables which are obtained by integrating the model in the required parameter range.}
\label{fig:tau40_Q1_bif}
\end{figure}

\begin{figure}
\begin{center}
\includegraphics[height=!, width=\textwidth]{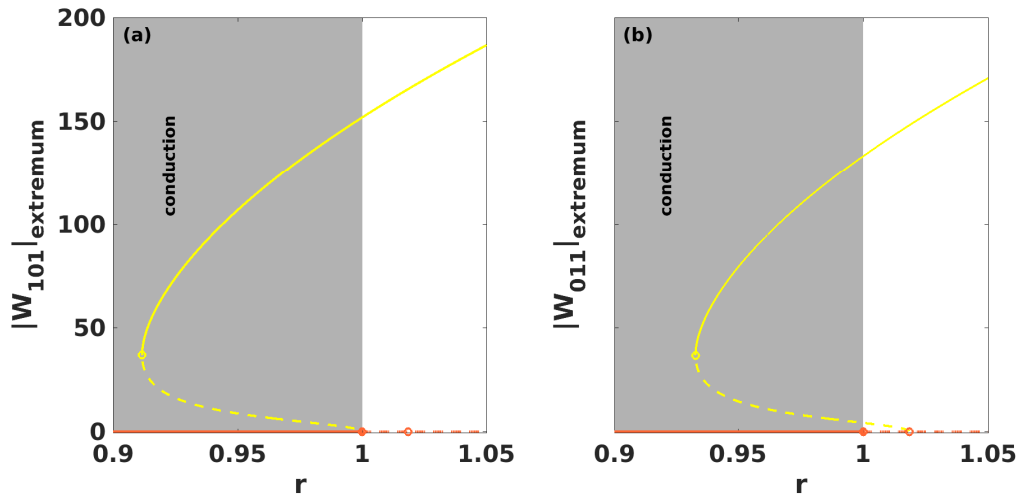}
\end{center}
\caption{Figure illustrates the bifurcations scenario at the onset of convection ($r = 1$) for the parameter set $\mathrm{Ta} = 50$, $\mathrm{Q} = 1$ and $\mathrm{Pr} = 0.025$. Two different roll solutions appear at the onset of convection in this case. Figure (a) shows bifurcation diagram where extremum values of $W_{101}$ is plotted in the vertical axis while extremum values of $W_{011}$ is plotted in figure (b).}
\label{fig:tau50_Q1_double_roll}
\end{figure}

Bifurcation structures also change as the value of $\mathrm{Ta}$ is varied for a fixed value of $\mathrm{Q}$. To illustrate those changes we prepare two separate bifurcation diagrams corresponding to $\mathrm{Ta} = 40$ and $\mathrm{Q} = 1$. The bifurcation diagrams are shown in figure~\ref{fig:tau40_Q1_bif}. We use the same line style and color coding in both figures as we have used in the previous bifurcation diagrams. From both figures we notice that all the bifurcations are originated from two different 2D rolls solutions as we have seen in the previous diagrams, not being connected by the symmetry. The stable conduction state bifurcates at $r = 1$ via a subcritical pitchfork bifurcation becomes unstable after that and gives birth to an unstable 2D roll branch for which $W_{101} \neq 0$ and $W_{011} = 0$ (yellow curve in figure~\ref{fig:tau40_Q1_bif}(a)). This unstable 2D roll branch moves backward initially, goes through a saddle-node bifurcation at $r = 0.9263$, becomes stable and starts to move forward. However, the stable 2D roll branch again becomes unstable at $r = 0.9556$ before coming out of the conduction region and a stable CR branch (blue curve) is originated there. Another 2D roll branch for which $W_{101} = 0$ and $W_{011} \neq 0$ is shown with yellow curve in figure~\ref{fig:tau40_Q1_bif}(b), shows similar behavior like the previous roll branch. It originates from a branch point of the unstable conduction state, moves backward until a saddle-node bifurcation point appears on it at $r = 0.9479$, becomes stable and continues to move forward after that. Later, at $r = 0.9754$ a stable CR$'$ branch (black curve) originates from a branch point of the 2D roll branch and the 2D roll branch again becomes unstable there. 
The stable CR branch undergoes a supercritical Hopf bifurcation at $r = 0.9635$, becomes unstable and stable limit cycle appears. The extrema of $W_{101}$ and $W_{011}$ corresponding to that limit cycles are shown with sea-green curves in both figures. However, the limit cycle exists inside the conduction region only. The unstable CR branch then undergoes through an inverse Hopf bifurcation at $r = 1.1909$ as we increase the value of $r$ and becomes stable. Due to this Hopf bifurcation, stable limit cycle appears and continues to exist for lower values of $r$. The extrema of the limit cycle are represented with red curves in both diagrams. The stable CR$'$ branch also undergoes through a Hopf bifurcation at $r = 0.9831$ and a stable limit cycle is produced there. This limit cycle also stays inside the conduction region only and extrema of the limit cycle are displayed with deep green curves in both figures. The CR$'$ branch continues to exist unstable throughout the parameter region after that Hopf bifurcation before it turns and meets a branch point located on 2D roll branch. As a result, the dynamics related to CR$'$ branch, which we have noticed for lower values of $\mathrm{Ta}$, disappears here. The limit cycle which appears through a Hopf bifurcation at $r = 1.1909$, becomes homoclinic at $r = 1.1076$ and we observe imperfect gluing bifurcation immediately after that in the range $1.0615 \leq r \leq 1.1077$. Simpler glued limit cycles appear followed by complicated glued limit cycles as we decrease the value of $r$ gradually. Further decrease in the value of $r$ gives complicated glued limit cycles again before we observe the ungluing bifurcation of the limit cycles near $r = 1.0614$. As a result of this ungluing bifurcation, we get the limit cycle in the rest of the range of $r$ till to the onset of convection. The extrema corresponding to unglued limit cycle are shown with brown dots in both figures while green dots represent extrema of gluing solutions. 
\begin{figure}
\begin{center}
\includegraphics[height=!, width=\textwidth]{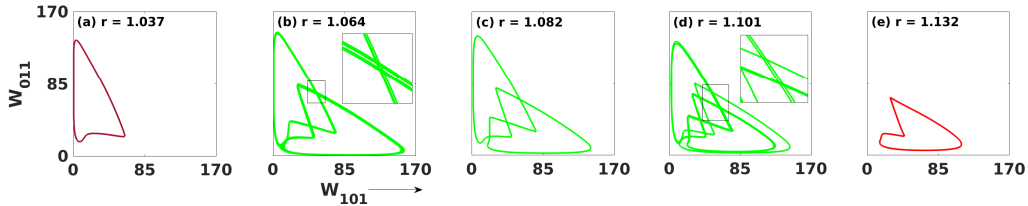}
\end{center}
\caption{Projections of phase space trajectories on $W_{101}-W_{011}$ plane is shown for $\mathrm{Ta} = 40$, for $\mathrm{Q} = 1$ corresponding to different values of $r$.}
\label{fig:tau40_Q1_phase_plot}
\end{figure}
Figure~\ref{fig:tau40_Q1_phase_plot} illustrates the imperfect gluing and ungluing phenomena corresponding to $\mathrm{Ta} = 40$ and $\mathrm{Q} = 1$ where, projections of phase space trajectories on $W_{101}-W_{011}$ plane are shown for different values of $r$. Further increment in the value of $\mathrm{Ta}$ for fixed value of $\mathrm{Q}$ expands the distance between saddle-node bifurcation point and the branch point from where the cross-roll branch originates in both 2D roll branches. As a result two different 2D roll branches exist at the onset of convection corresponding to two different set of initial conditions. This scenario is shown in figure~\ref{fig:tau50_Q1_double_roll} for $\mathrm{Ta} = 50$ and $\mathrm{Q} = 1$, where two different rolls solutions appear at the onset of convection. Note that these two rolls solutions are not connected by system symmetry but exist independently at the onset. We observe similar results from DNS also. 

\subsection{Imperfect gluing bifurcation}
Symmetries play a crucial role in determining the behavior of nonlinear physical systems. A well-known example is supercritical pitchfork bifurcation where a pair of stable solutions appear as we vary a parameter. As a result, the original system symmetry (reflection symmetry) is broken by these new states and these new states become the symmetric images of each other. However, it is very hard to achieve perfect symmetry in any physical system. Therefore, bifurcations we observe in physical systems may become imperfect. This can be verified easily by adding a constant term in the normal form of the supercritical pitchfork bifurcation which breaks the reflection symmetry of the system and acts like an imperfection parameter. Due to this change, the bifurcation structure becomes disjoint in which one branch varies monotonically with the parameter and another one appears through a saddle-node bifurcation. Here we are concerned with global bifurcations specially homoclinic gluing bifurcations in presence of imperfection. Imperfect homoclinic gluing bifurcation has been the topic of interest to researchers for a long time and has been reported earlier in various problems including Taylor-Couette flow~\citep[]{marques:2002}, dynamically coupled extended flow~\citep[]{abshagen:PRL_2001} and electronic circuits~\citep[]{glendinning:2001}. In the case of rotating convection when the external horizontal magnetic field is absent, the system preserves the $x\rightleftharpoons y$ symmetry, two stable limit cycles, symmetric images of each other become homoclinic to a saddle (SQR) at $r = r_g$ as the value of $r$ is decreased and symmetric glued limit cycles appear after that for subsequent lower values of $r$~\citep[]{maity:EPL_2013}.  
\begin{figure}
\begin{center}
\includegraphics[height=!, width=\textwidth]{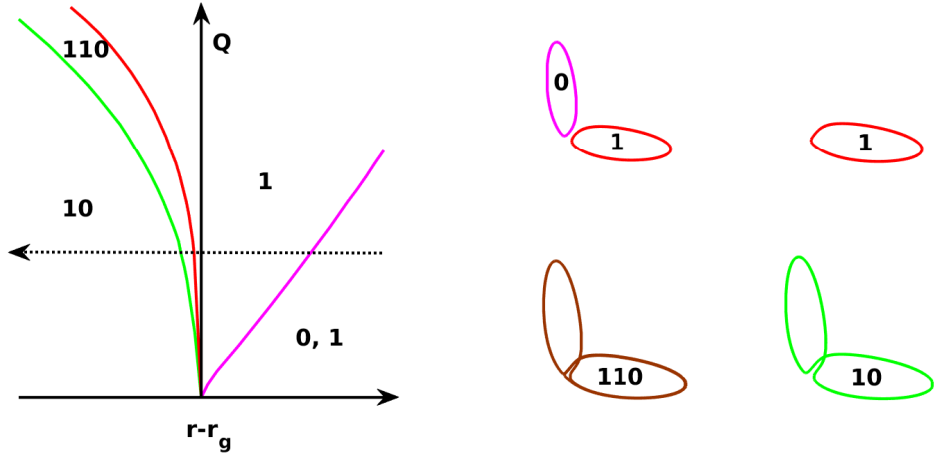}
\end{center}
\caption{The figure in left panel shows two parameter ($r$ vs $\mathrm{Q}$) plane corresponding to imperfect gluing bifurcation for $\mathrm{Ta} = 20$ for $\mathrm{Pr} = 0.025$. We get different type of periodic orbits as we move from right hand side to left hand side following the arrow for a fixed value of $\mathrm{Q}$. The projection of different type of periodic orbits on $W_{101}-W_{011}$ plane are displayed in the right panel of the figure corresponding to different values of $r$.}
\label{fig:imperfect_gluing}
\end{figure}
However, the presence of an external magnetic field in the horizontal direction in RMC breaks the $x\rightleftharpoons y$ symmetry of the system, which gives rise the imperfection. Therefore, $\mathrm{Q}$ is the imperfection parameter in this case and its value is the measure of imperfection with $\mathrm{Q} = 0$ refers to a system with perfect symmetry. As a result, two different limit cycles continue to exist independently not being connected by the symmetry. Later these two limit cycles become homoclinic to SQR at two separate points and asymmetric glued limit cycles along with some complicated periodic orbits are observed after that. To describe the effect of imperfection on homoclinic gluing bifurcation we prepare a two-parameter bifurcation diagram for $\mathrm{Ta} = 20$, for $\mathrm{Pr} = 0.025$ (see left panel of figure~\ref{fig:imperfect_gluing}). In the diagram $r-r_g$ is plotted against the imperfection parameter $\mathrm{Q}$. The origin in the figure $(r_g, 0)$ is the point at which, two symmetrically connected limit cycles become homoclinic and form symmetric glued limit cycles thereafter as we move to left from right, in the absence of imperfection. In presence of $\mathrm{Q}$ when two limit cycles exist independently we encoded them with `$0$' and `$1$'. Note that `$0$' and `$1$' here represent the OCR-II$'$ and OCR-II solutions respectively, while `$10$' represents the OCR-I solutions. Also when $\mathrm{Q} = 0$, `$0$' and `$1$' are the symmetric images of each other and `$10$' is symmetric. From the figure we see that three curves, viz. pink, red and green divide the two parameter plane into four regions where qualitatively different solutions exist. To explore these solutions we traverse the two parameter plane from right side to left for a fixed value of $\mathrm{Q}$ ($\mathrm{Q} = 0.2$) following the arrow. We start from the region which is placed in the right side of the pink curve where two limit cycles exist independently, not connected by the $x\rightleftharpoons y$ symmetry. Projection of these two limit cycles on $W_{101}-W_{011}$ plane is shown on the right panel of the figure with code `$0$' and `$1$'. Now as we move with the arrow, the arrow intersects the pink curve at a point where limit cycle `$0$' becomes homoclinic and destroys there. Moving further we reach the region which lies between the pink and red curves. Only limit cycle `$1$' exists in this region (see figure~\ref{fig:imperfect_gluing}). This limit cycle goes through a homoclinic bifurcation at the point where the arrow cuts the red curve. Soon after, this homoclinic bifurcation, we observe asymmetric glued limit cycles in the region lies in the left side of the red curve. However, a complicated glued limit cycle which encircles the `$1$' limit cycle exists in between green and red curves. Simpler glued limit cycles exist in the region located in the left side of the green curve. Phase portraits of the different glued limit cycle are shown in the right panel of figure~\ref{fig:imperfect_gluing}. 
\begin{figure}
\begin{center}
\includegraphics[height=!, width=\textwidth]{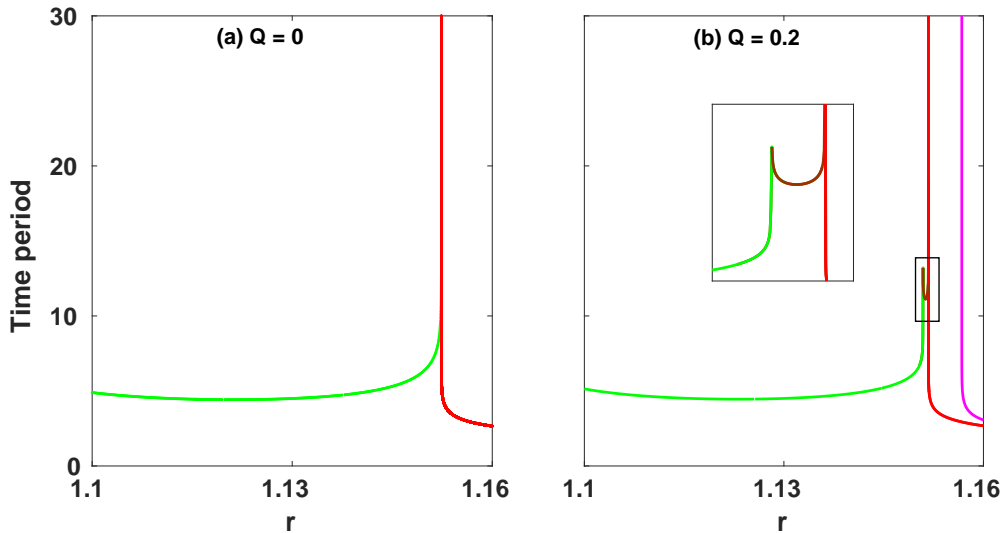}
\end{center}
\caption{Figure represents variation of time period of oscillation corresponding to different periodic solutions for $\mathrm{Ta} = 20$, for $\mathrm{Pr} = 0.025$ and different values of $\mathrm{Q}$. Figure (a) displays the variation of time period for different solutions for $\mathrm{Q} = 0$, while fig (b) shows the same corresponding to $\mathrm{Q} = 0.2$. The red curve in figure (a) exhibits the variation of time period of two symmetrically connected limit cycles near homoclinic bifurcation point, while the green curve shows the time period of glued solution. In figure (b) the pink and red curves display variation of time period corresponding to the limit cycles coded as `$0$' and `$1$' respectively near the homoclinic bifurcation points. The green curve exhibits the time period of gluing solutions as usual.}
\label{fig:imperfect_time_period}
\end{figure}
Variation of time period corresponding to different periodic orbits are shown in figure~\ref{fig:imperfect_time_period}. Figure~\ref{fig:imperfect_time_period}(a) displays variation of time period of periodic orbits near homoclinic bifurcation point corresponding to $\mathrm{Q} = 0$ i.e., in presence of perfect symmetry. In the figure, we see that as the limit cycles, symmetric images of each other approach to the bifurcation point, time period of their oscillation becomes very large (red curve). While glued limit cycles are observed right after the bifurcation point whose time period of oscillation is shown with green curve in the figure. Presence of imperfection changes the scenario shown in figure~\ref{fig:imperfect_time_period}(b). From figure, we see that due to the symmetry breaking, time period of limit cycle coded with `$0$' diverges first (pink curve). Then the time period of limit cycle `$1$' diverges (red curve). Also time period of the gluing solution (green curve) diverges and a complicated periodic orbit coded `$110$' appears. Note that, this imperfect bifurcation scenario exists for small value of $\mathrm{Q}$ and vanishes for higher values of $\mathrm{Q}$. 

\subsection{Oscillatory instability of 2D rolls}
\begin{figure}
\begin{center}
\includegraphics[height=!, width=\textwidth]{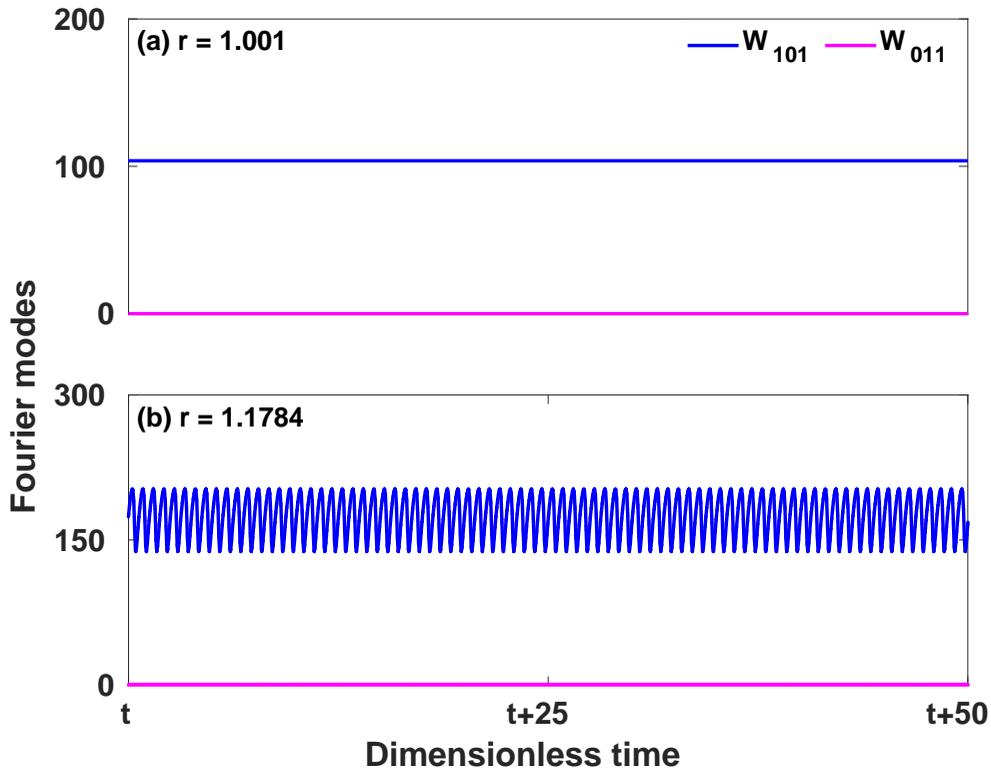}
\end{center}
\caption{Temporal variation of the Fourier modes $W_{101}$ (blue curve) and $W_{011}$ (pink curve) computed from DNS corresponding to $r = 1.001$ and $r = 1.1784$ for $\mathrm{Ta} = 20$, $\mathrm{Q} = 50$ and $\mathrm{Pr} = 0.025$ are shown in the figure.}
\label{fig:busse_instability_dns}
\end{figure}

\begin{figure}
\begin{center}
\includegraphics[height=!, width=\textwidth]{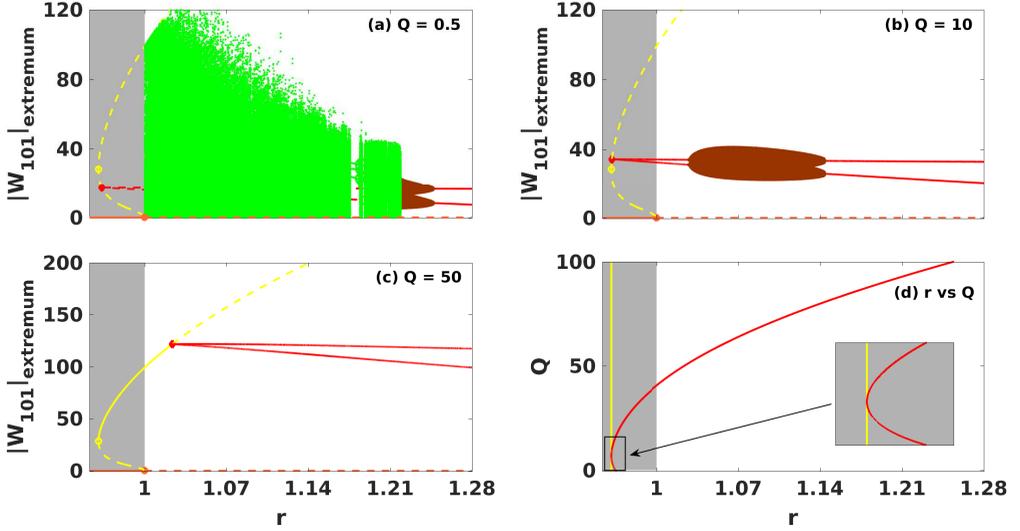}
\end{center}
\caption{Bifurcation diagrams constructed using $27$-mode model for $\mathrm{Ta} = 20$, $\mathrm{Pr} = 0.025$ and three different values of $\mathrm{Q}$ together with a two-parameter diagram are shown in the figure. Fig (a) displays the bifurcation diagram corresponding to $\mathrm{Q} = 0.5$, while fig (b) exhibits the bifurcation diagram for $\mathrm{Q} = 10$ and bifurcation diagram corresponding to $\mathrm{Q} = 50$ is shown in fig (c). Same color coding is used in all three bifurcation diagrams. The stable and unstable solutions are represented by solid and dashed curve respectively. Orange, yellow and red curves display the conduction state, steady 2D rolls along y-axis and periodic wavy rolls respectively. Green and brown dots in the figures are chaotic and quasiperiodic wavy rolls. The empty circle and filled diamond are the limit-point and Hopf bifurcation point. Shaded gray region is the conduction region. Continuation of the limit-point and Hpof bifurcation point over $r-\mathrm{Q}$ plane is shown in fig (d). The yellow curve is the continuation of limit-point and continuation of the Hopf bifurcation point is represented by the red curve. Near $\mathrm{Q} \approx 6.7$  the Hopf bifurcation point crosses the limit-point which is shown in the inset.}
\label{fig:busse_instability}
\end{figure}

We identify a different set of initial conditions during performing DNS using random initial conditions for which we observe steady 2D rolls ($W_{101} \neq 0$, $W_{011} = 0$) at the onset of convection and time dependent solutions for which $W_{101} \neq 0$, $W_{011} = 0$ for higher values of $r$. The time series corresponding to these solutions are shown in figure~\ref{fig:busse_instability_dns}. These time dependent solutions are qualitatively different from the time dependent solutions we have seen in previous section, where both modes $W_{101}$ and $W_{011}$ have nonzero values. To understand the origin of these solutions and associated bifurcation structures we prepare a low-dimensional model which is different from the previous one. We start with the vertical velocity mode $W_{101}$, vertical vorticity mode $Z_{010}$ and temperature modes $T_{101}$ and $T_{002}$. We follow same low-dimensional modeling technique which we have used previously to construct $39$-mode model. Following the procedure, we find seven vertical velocity modes: $W_{101}$, $W_{121}$, $W_{202}$, $W_{022}$, $W_{301}$, $W_{103}$, $W_{111}$, twelve vertical vorticity modes: $Z_{101}$, $Z_{121}$, $Z_{202}$, $Z_{022}$, $Z_{103}$, $Z_{301}$, $Z_{010}$, $Z_{111}$, $Z_{012}$, $Z_{210}$, $Z_{020}$, $Z_{200}$, and eight temperature modes: $T_{101}$, $T_{121}$, $T_{202}$, $T_{022}$, $T_{301}$, $T_{103}$, $T_{111}$, $T_{002}$, which are the most energetic modes. Projecting the hydrodynamic equations on these modes we get a set of $27$ coupled nonlinear ordinary differential equations which is the low-dimensional model in this case. 
We perform bifurcation analysis of this $27$-mode model to unfold the bifurcation scenario and prepare three qualitatively different bifurcation diagrams along with a two-parameter diagram (see figure~\ref{fig:busse_instability}) corresponding to $\mathrm{Pr} = 0.025$. Bifurcation diagram corresponding to $\mathrm{Ta} = 20$ and $\mathrm{Q} = 0.5$ constructed using this model is shown in figure~\ref{fig:busse_instability}(a). In the figure the stable conduction state is displayed with the solid orange curve, which becomes unstable at $r = 1$ through a subcritical pitchfork bifurcation and an unstable 2D roll branch is generated there. The unstable conduction solution continues to exist for higher values of $r$ and is represented by the dashed orange curve in the bifurcation diagram. The unstable 2D roll branch moves backward at the beginning and starts to move forward after a limit-point (yellow open circle) appears on it at $r = 0.9604$ and comes out of the conduction region (shaded grey region) again at $r = 1$. This unstable 2D roll branch undergoes through a Hopf bifurcation at $r = 0.9634$ (filled red diamond) before the limit-point appears on it and an unstable limit cycle is originated from there. The extrema of $W_{101}$ for this unstable limit cycle are shown with dashed red curves. This limit cycle undergoes an inverse Neimark-Sacker (NS) bifurcation at $r = 1.2481$ and becomes stable. The stable limit cycle is represented by solid red curves. Quasiperiodic solutions appear for lower values of $r$ as a result of this inverse NS bifurcation, which are shown with brown dots. For further lower values of $r$ till onset we observe chaotic solutions via intermediate phase locked states. The green dots exhibit the chaotic solutions in the figure. The origin of these time-dependent solutions in this case is the 2D rolls solution, similar to the one reported by Busse~\citep[]{busse:JFM_1972}. Physically these time-dependent solutions represent wavy roll patterns. Next, we prepare the bifurcation diagram for $\mathrm{Ta} = 20$ and $\mathrm{Q} = 10$ (see figure~\ref{fig:busse_instability}(b)) which is qualitatively different from the previous one. The unstable 2D roll branch in this case goes through a saddle-node bifurcation at $r = 0.9604$ first and become stable. The stable 2D roll branch then goes through a Hopf bifurcation at $r = 0.9610$ and a stable limit cycle is originated from there. The extremum values of $W_{101}$ corresponding to these limit cycles are shown with solid red curves in the figure. This limit cycle goes through an NS bifurcation at $r = 1.0265$, become unstable and quasiperiodic solutions (brown dots) appear for higher values of $r$. At $r = 1.1451$ the unstable limit cycle becomes again stable through an inverse NS bifurcation and continues to exist as we increase the value of $r$ further. Note that we do not observe any chaotic flow in this case. As we increase the value of $\mathrm{Q}$ further we see more qualitative changes in bifurcation structure. To illustrate those changes we prepare another bifurcation diagram for $\mathrm{Ta} = 20$ and $\mathrm{Q} = 50$ (see figure~\ref{fig:busse_instability}(c)). From the figure we see that the unstable 2D roll branch which moves backward at starting, goes through a saddle-node bifurcation, becomes stable, turns around, starts to move forward and comes out of the conduction region. As a result we notice stable steady 2D rolls solution at the onset of convection. This stable 2D roll branch goes through a Hopf bifurcation at $r = 1.0236$ as we increase the value of $r$. Stable limit cycle appears due to this Hopf bifurcation which continues to exist for higher values of $r$. Solid red curves in the figure~\ref{fig:busse_instability}(c) exhibit the extrema of the $W_{101}$ corresponding to that limit cycle. Chaotic and quasiperiodic solutions are absent in this case. Noting all three bifurcation diagrams we find that chaotic solutions exist when the value of $\mathrm{Q}$ is very small for a fixed value of $\mathrm{Ta}$ i.e., when the strength of the external magnetic field is very low for a fixed rotation rate. As we increase the value of $\mathrm{Q}$ i.e., the strength of the external magnetic field the chaotic solutions get suppressed and quasiperiodic solutions develop. Further increment in the value of $\mathrm{Q}$ eliminates the quasiperiodic solutions and only periodic solutions sustain. The onset of convection also changes depending on the value of $\mathrm{Q}$. Chaotic solutions, which appear at the onset of convection for small values of $\mathrm{Q}$, get suppressed as the value of $\mathrm{Q}$ is increased. As a result, periodic solutions develop at the onset, which also get suppressed as we raise the value of $\mathrm{Q}$ further and stable 2D rolls solutions appear. Once 2D rolls solutions appear at the onset for a particular value of $\mathrm{Q}$ the onset remains unaltered for subsequent higher values of $\mathrm{Q}$. Therefore increment in the strength of the external magnetic field gradually suppresses the chaotic flows when the rotation rate is fixed. This is an interesting phenomenon which has been reported earlier by~\citep{pal:2012} in presence of an external horizontal magnetic 
field. The appearance of limit-point and Hopf bifurcation point play an important roll in determining the stability of 2D roll branch. From figure~\ref{fig:busse_instability}, we see that the Hopf bifurcation point appears before limit-point when the value of $\mathrm{Q}$ is sufficiently small. Therefore 2D roll branch remains unstable throughout the considered range of $r$. Now as we increase the value of $\mathrm{Q}$ slowly the Hopf bifurcation point starts to move towards the limit-point and crosses it near $\mathrm{Q} \approx 6.7$. As a result, the 2D roll branch now goes through a saddle-node bifurcation first instead of a Hopf bifurcation and becomes stable for subsequent higher values of $\mathrm{Q}$. The continuation of the limit-point and the Hopf bifurcation point is shown in figure~\ref{fig:busse_instability}(d). The yellow curve in the figure is the continuation of the limit-point over two-parameter plane i.e., $\mathrm{Q}$ against $r$. While the red curve represents the continuation of Hopf bifurcation point in $r-\mathrm{Q}$ plane. The scenario where the Hopf bifurcation point crosses the limit-point near $\mathrm{Q} \approx 6.7$ is shown in the inset. However, stable 2D rolls appears at the onset of convection as soon as the red curve in figure comes out of the conduction region near $\mathrm{Q} \approx 40.7$ and continues to exist for higher values of $\mathrm{Q}$. The results from DNS show good qualitative agreement with the model results in this case also. 

\section{Conclusions}
In this paper, we have investigated convective instabilities, bifurcations and patterns near the onset of Rayleigh-B\'{e}nard convection of electrically conducting fluids in presence of rotation about vertical axis as well as external uniform horizontal magnetic field. The investigation is carried out by performing three dimensional (3D) direct numerical simulations (DNS) of the governing equations and theoretical analysis of low dimensional models constructed from DNS data. We have considered the values of Chandrasekhar number ($\mathrm{Q}$) and Taylor number ($\mathrm{Ta}$) in the ranges $0\leq \mathrm{Q} \leq 50$ and $0\leq \mathrm{Ta}\leq 100$. 

Primary DNS of the system show some interesting results including a plethora of stationary and time dependent patterns just at the onset of convection for different sets of values of $\mathrm{Q}$ and $\mathrm{Ta}$ in the above mentioned ranges.  Moreover, for relatively higher values of $\mathrm{Q}$, a substantial enhancement of the convective heat transfer is observed. 

These prompted us to understand the origin of these flow patterns as well as the reason for notable increase of heat transfer near the onset of convection. Low dimensional modeling of the system from the DNS data helped to understand the bifurcation structure and the origin of different flow patterns near the onset of convection quite convincingly in the considered ranges of $\mathrm{Q}$ and $\mathrm{Ta}$. A very rich bifurcation structure has been discovered near the onset of convection involving a range of local as well as global bifurcations. 

Performance of DNS as well as low dimensional modeling of the system establishes a first order transition to convection at the onset in a wide range of the parameter space which is responsible for the enhancement of the heat transfer at the onset. The first order transition to convection is found to be associated with a subcritical pitchfork bifurcation. Interestingly, we have presented a convenient $3$-mode model which captures the phenomenon of first order transition to convection in a wide range of the parameter space quite accurately.\\

\noindent {\bf Acknowledgements}\\
P.P. acknowledges support from Science and Engineering Research Board (Department of Science and Technology, India) (Grant No. EMR/2015/001680). M.G. is supported by INSPIRE programme of DST, India (Code: IF150261). Authors thank Ankan Banerjee for fruitful
comments.

\bibliographystyle{jfm}
\bibliography{manojit_rmc}

\end{document}